\documentclass[aip,jcp,reprint]{revtex4-1} 
\usepackage{graphicx}
\usepackage{enumitem}

\usepackage{mathtools}
\usepackage{graphicx}   
\usepackage{color}
\usepackage{ulem}
\usepackage{caption}
\usepackage{subcaption}
\usepackage{comment}
\usepackage{bm}

\usepackage{algorithm}
\usepackage{algpseudocode}
\usepackage{algorithmicx}

\usepackage{array}
\usepackage{booktabs}
\usepackage{multirow}

\usepackage{amsmath}

\usepackage{booktabs}

\begin{document}
\title{State Predictive Information Bottleneck}

\author{Dedi Wang}
\affiliation{Biophysics Program and Institute for Physical Science and Technology, 
University of Maryland, College Park 20742, USA.}

\author{Pratyush Tiwary\footnote{Corresponding author.}}
 
\email{ptiwary@umd.edu}
\affiliation{Department of Chemistry and Biochemistry and Institute for Physical Science and Technology, University of Maryland, College Park 20742, USA.}

\date{\today}
	
\begin{abstract}
The ability to make sense of the massive amounts of high-dimensional data generated from  molecular dynamics (MD) simulations is heavily dependent on the knowledge of a low dimensional manifold (parameterized by a reaction coordinate or RC) that typically distinguishes between relevant metastable states and which captures the relevant slow dynamics of interest. Methods based on machine learning and artificial intelligence have been proposed over the years to deal with learning such low-dimensional manifolds, but they are often criticized for a disconnect from more traditional and physically interpretable  approaches. To deal with such concerns, in this work, we propose a deep learning based State Predictive Information Bottleneck (SPIB) approach to learn the RC from high dimensional molecular simulation trajectories. We demonstrate analytically and numerically how the RC learnt in this approach is deeply connected to the committor in chemical physics, and can be used to accurately identify transition states. A crucial hyperparameter in this approach is the time-delay, or how far into the future the algorithm should make predictions about. Through careful comparisons for benchmark systems, we demonstrate that this hyperparameter choice gives useful control over how coarse-grained we want the metastable state classification of the system to be. We thus believe that this work represents a step forward in systematic application of deep learning based ideas to molecular simulations in a way that bridges the gap between artificial intelligence and traditional chemical physics.
\end{abstract}

\maketitle
\section{Introduction}

Rapid advances in computational power have made molecular dynamics (MD) a powerful tool for studying systems in biophysics, chemical physics and beyond. However, there are still at least two open questions in this area: first, how to make use of the deluge of data generated from MD simulation understandable for a human; second, how to further extend timescales that can be reached in MD. The unifying aspect to overcoming both these difficulties is to efficiently uncover a low dimensional manifold (parameterized by a reaction coordinate or RC) on which the dynamics of the system can be projected.\cite{ML_review}

Over the past decades, various approaches have been developed to learn the RC from trajectory data. It has been argued that for given two states, the committor, defined next, is a perfect candidate for the RC as it provides a quantitative description of the dynamics along a trajectory.\cite{committor_aladip,committor_review} Let A and B denote the reactant and product states, then the committor probability $p_B(x)$ is defined as the probability of the trajectories that reach state B prior to the state A from a conformation $X$. Through the analysis of the committor distribution, much insight has been obtained in a variety of phenomena ranging from ion solvation to biomolecular isomerization,\cite{committor_aladip,committor_ion,Pluharova2016,Roy2016} Transition path sampling (TPS), which focuses on sampling the pathways connecting metastable states, is a powerful tool to analyze the committor.\cite{TPS,committor_review,HummerPNAS2005} Based on it, some physically meaningful RC can then be identified through a generic algorithm\cite{Ma2005} or a likelihood maximization approach\cite{peters2006obtaining,peters2016}. However, these methods always heavily depend on human intuition to generate the trial coordinates.\cite{peters2016}

Another approach to obtaining the RC is to learn the relevant slow modes of dynamics. Coifman, Kevrekidis, Clementi and others first used diffusion map to determine collective reaction coordinates for macromolecular dynamics.\cite{diffusion_map_and_RC,Coifman2008,Rohrdanz2011} Thereafter, Noé and coworkers proposed the variational approach to conformation dynamics (VAC) and combined it with the time-lagged independent component analysis (TICA) to identify the optimal ``slow subspace" from a large set of prior order parameters.\cite{VAC,TICA} More recently a generalized version called VAMPnets was developed by the same group leveraging the power of neural networks.\cite{VAMPnets} In a similar spirit, the SGOOP method by Tiwary and Berne used an iterative approach to find RC through a maximum path entropy framework.\cite{SGOOP} Though all these slow-mode based methods are highly interpretable, the optimization can usually be difficult unless some simplifications are made. For instance, in TICA and SGOOP these simplifications could include learning the RC as linear combinations of pre-selected order parameters.\cite{TICA,SGOOP}

Besides these two physics-based approaches, other statistical approaches have also been developed to learn RC through a more flexible framework, such as VDE\cite{VDE} and RAVE\cite{RAVE,pRAVE}. In the RAVE approach for instance, the RC is interpreted as a bottleneck or a low dimensional space that predicts the most important features of the simulated trajectories. Such a RC can then be learned by making a trade-off between prediction and model complexity through an objective function. Typically a variational Bayesian approach is employed to allow these methods to parameterize the objective function using a neural network and achieve highly efficient training.\cite{VAE,variational_IB} Arguably these methods can be less expensive than physics-based approaches, but they may also be uninterpretable and therefore unreliable since the relationship between the statistics-based RC and the traditional physics-based RC is still unclear.

In this work, we develop a State Predictive Information Bottleneck (SPIB) framework that allows us to efficiently and accurately learn a RC from MD trajectories. Most importantly, we demonstrate rigorously how the slow bottleneck variable learnt in RAVE and related deep learning based methods can qualify as a good RC with the same attributes as expected from the committor. Similar to RAVE\cite{RAVE,pRAVE}, we also assume that RC should carry only the minimal information of the past to still be able to reliably predict the future state of the system.  The key feature that makes SPIB stand out is that a discrete-state representation of this system is learned on-the-fly during the training process and guides our RC to focus only on the motion related to the state-to-state transitions. We show analytically and numerically that the RC learned by our algorithm is related to the committor, and demonstrate that it can capture the important information from the trajectory to identify the correct transition state. Moreover, we demonstrate how our algorithm can automatically figure out the metastable states in a complex system and generate an accurate but still highly understandable description of their inter-conversion dynamics. Given these promising properties, we believe our algorithm can be a powerful tool to analyze generic complex systems.

\section{Method}
\label{sec:method}

\subsection{Information Bottleneck}
\label{sec:information_bottleneck}
The Information Bottleneck (IB) principle provides a general framework to learn a concise representation $\bm{z}$ of an input source $\bm{X}$ that is maximally informative  about some target $\bm{y}$.\cite{IB,variational_IB} Here typically the representation $\bm{z}$ has much smaller dimensionality than the source $\bm{X}$, while the target $\bm{y}$ can be of low or high dimensionality depending on the task at hand. The IB principle  postulates that the desired representation $\bm{z}$ should use minimal information from the input $\bm{X}$ to predict the target $\bm{y}$. Mathematically, such a learning process can be formulated as maximizing the objective function: 
\begin{equation} 
\label{eq:IB_obj}
\mathcal{L_{IB}}\equiv I(\bm{z},\bm{y})  - \beta I(\bm{X},\bm{z})
\end{equation} 

Here, the function $I(x,y)\equiv  \int dx dy\ p(x,y)\log \frac{p(x,y)}{p(x)p(y)}$ denotes the mutual information between any two random variables. The trade-off between the prediction capacity $I(\bm{z},\bm{y})$ and model complexity $I(\bm{X},\bm{z})$ is controlled by the Lagrange multiplier $\beta\in[0,\infty)$. Unfortunately, the direct optimization of the information bottleneck shown in Eq. \ref{eq:IB_obj} is impractical as the calculation of mutual information in general is computationally expensive. \cite{variational_IB,pRAVE} Thus, following Ref. \onlinecite{variational_IB}, we can obtain a variational lower bound on the original objective function from Eq. \ref{eq:IB_obj}:

\begin{equation} 
\begin{aligned} 
\label{eq:variational_IB_obj}
&\mathcal{L_{IB}}\ge -\underbrace{\frac{1}{N}\sum_{n=1}^N\int d\bm{z}\Bigl[ -p(\bm{z}|\bm{X}^n)\log q(\bm{y}^n|\bm{z})\Bigr] }_{distortion} \\
&-\beta\ \underbrace{\frac{1}{N}\sum_{n=1}^N\int d\bm{z}\Bigl[p(\bm{z}|\bm{X}^n)\log \frac{p(\bm{z}|\bm{X}^n)}{r(\bm{z})} \Bigr]}_{rate}+H(\bm{y})=\mathcal{L}
\end{aligned} 
\end{equation}
where $q(\bm{y}|\bm{z})$ and $r(\bm{z})$ are variational approximations to the true probability distributions $p(\bm{y}|\bm{z})$ and $p(\bm{z})$ respectively. Notice that the entropy of the targets $H(\bm{y})\equiv-\int dy\ p(\bm{y})\log p(\bm{y})$ in Eq. \ref{eq:variational_IB_obj} is independent of the optimization process and hence can be ignored. From a coding theory perspective,\cite{coding_theory} as $\bm{z}$ can be interpreted as a latent representation or a code, we usually refer to $p(\bm{z}|\bm{X})$ as a probabilistic encoder, and $q(\bm{y}|\bm{z})$ as a probabilistic decoder. Interestingly, one can easily obtain from Eq. \ref{eq:variational_IB_obj} the objective function used in variational autoencoders by assuming $\beta=1$ and requiring the representation $\bm{z}$ to reconstruct the input $\bm{X}$ instead of predicting a target $\bm{y}$.\cite{VAE,variational_IB} Based on rate-distortion theory,\cite{RD_theory,alemi2018an} the first term in Eq. \ref{eq:variational_IB_obj} can be interpreted as the distortion, which measures the ability of our representation to predict the desired target, while the second term can be interpreted as the rate, which measures the number of bits per data sample to be transmitted. Thus, maximizing $\mathcal{L}$ can also be viewed as the problem of  determining the minimal number of bits, as measured by the rate, that should be communicated from a source through a channel so that the receiver can reconstruct the original signal without exceeding a desired value of the distortion. 

There are many possible choices for the encoder $p(\bm{z}|\bm{X})$, the decoder $q(\bm{y}|\bm{z})$ and the approximate prior $r(\bm{z})$, depending on the particular application domain. We point out here that all these three probability distributions can depend collectively on some model parameters $\theta$, which are learned during the training process. Therefore, in the following sections, we will add a subscript $\theta$ to all these three distributions $\left\{ p_{\theta}(\bm{z}|\bm{X}), q_{\theta}(\bm{y}|\bm{z}), r_{\theta}(\bm{z}) \right\}$.

\subsection{State Predictive Information Bottleneck}
\label{sec:state_predictive_information_bottleneck}

The generic IB framework introduced in Sec. \ref{sec:information_bottleneck} leaves ample scope for the specific flavor of implementation in many different ways, as for instance we demonstrated in our past publications through the RAVE family of methods,\cite{RAVE,pRAVE} and it has been discussed more generally in Ref. \onlinecite{alemi2018therml}. Based on the general IB framework, in this section we advance our RAVE family of methods with a State Predictive Information Bottleneck (SPIB) framework. Similar to existing RAVE formulations here as well we aim to learn an accurate reaction coordinate (RC) for generic molecular systems, but
make RAVE significantly more robust in many aspects, and draw rigorous and useful connections between the past-future information bottleneck and the committor based definition of the RC in theoretical chemistry.\cite{committor_aladip,best2005reaction} However unlike RAVE, where the aim is to predict a time-delayed version of the entire input molecular configuration, here we set as  target $\bm{y}$ in Eq. \ref{eq:IB_obj} its future state, which is drawn from a dictionary of indices for possible metastable states. The target $\bm{y}$ is relatively much lower in dimensionality than the exact molecular configuration. In this way, we require our RC to only predict which state the system will stay in after a time delay $\Delta t$, instead of its exact configuration. Typically the number and location of such states are not available \textit{a priori} and our work makes it possible to estimate these robustly and on-the-fly, as we demonstrate in Sec. \ref{sec:iterative_retraining_algorithm}.

The main advantage of such a simplification of the prediction task is that only the motion related to the transitions between different states will be captured by the learnt RC,  while the fluctuations inside any metastable state will be ignored.
Thus, for a given unbiased trajectory $\{\bm{X}^1,\cdots,\bm{X}^{M+s}\}$ and its corresponding state labels $\{\bm{y}^1,\cdots,\bm{y}^{M+s}\}$ with large enough $M$, the objective function of SPIB can be formulated as:
\begin{equation} 
\begin{aligned} 
\label{eq:SPIB_obj}
\mathcal{L}\approx \frac{1}{M\cdot L}\sum_{n=1}^M\sum_{l=1}^L &\Bigl[\log q_{\theta}(\bm{y}^{n+s}|\bm{z}^{(n,l)}) \\
&-\beta \log \frac{p_{\theta}(\bm{z}^{(n,l)}|\bm{X}^n)}{r_{\theta}(\bm{z}^{(n,l)})} \Bigr]
\end{aligned} 
\end{equation}
where $\bm{z}^{(n,l)}$ is sampled from $p_{\theta}(\bm{z}|\bm{X}^n)$ and the time interval between $\bm{X}^n$ and $\bm{X}^{n+s}$ is the time delay $\Delta t$.  

In SPIB, the trajectory $\{\bm{X}^n\}$ is usually expressed in terms of many order parameters or features, while the state labels $\{\bm{y}^n\}$ are mutually exclusive and expressed in terms of one-hot vectors, i.e. a binary vector with a single high (1) bit and all the others low (0). To implement this we use a deep feed forward neural network with softmax outputs in our decoder $q_{\theta}(\bm{y}|\bm{z})$. 
\begin{equation} 
\label{eq:softmax_decoder}
\log q_{\theta}(\bm{y}^{n+s}|\bm{z}^n) = \sum_{i=1}^{D} y_i^{n+s}\log \mathcal{D}_i(\bm{z}^n;\theta)
\end{equation} 
where the state label $\bm{y}$ is a one-hot vector of $D$ dimensions and the decoder function $\bm{\mathcal{D}}$ is the $D$-dimensional softmax output of a neural network.

Given that we expect the learnt RC should demarcate between different metastable states, it is natural to assume a multi-modal distribution for the prior $r_{\theta}(\bm{z})$.
In our algorithm, we employ the variational mixture of posteriors prior (VampPrior) to obtain such a multi-modal prior distribution.\cite{VampPrior} Here, the approximate prior $r_{\theta}(\bm{z})$ is a weighted mixture of different posteriors $p_{\theta}(\bm{z}|\bm{X})$ with pseudo-inputs $\{\bm{u}^k\}_{k=1}^K$  in lieu of $\bm{X}$:
\begin{equation} 
\label{eq:vamp_prior}
r_{\theta}(\bm{z}) = \sum_{k=1}^{K}\omega_k\ p_{\theta}(\bm{z}|\bm{u}^k)
\end{equation} 
where $K$ is the number of pseudo-inputs, $\bm{u}^k$ is a vector which has the same dimension as input $\bm{X}$, and $\omega_k$ represents the weight of $p_{\theta}(\bm{z}|\bm{u}^k)$ under the constraint $\sum_{k}\omega_k=1$. The pseudo-inputs $\{\bm{u}^k\}$ and weights $\{\omega_k\}$ can be thought of the parameters of the prior, which are learned through backpropagation of the objective function (Eq. \ref{eq:SPIB_obj}). In principle, the number of pseudo-inputs should be equal to the number of metastable states in the system. In practical settings however for real-world applications to complex molecular systems, the number of metastable states is unknown \textit{a priori}. To deal with such cases, the simple and powerful solution is to choose a large enough $K$ making the prior more flexible.

Finally, for simplicity, we take the encoder $p_{\theta}(\bm{z}|\bm{X})$ in Eq. \ref{eq:vamp_prior} as a neural network with a multivariate Gaussian output:
\begin{equation} 
\label{eq:gaussian_encoder}
\log p_{\theta}(\bm{z}^n|\bm{X}^n) = \log \mathcal{N}(\bm{z}^n;\bm{\mu},\bm{\sigma} I)
\end{equation}
where the mean $\bm{\mu}$ and variance $\bm{\sigma}^2$ are outputs of a neural network whose input is $\bm{X}^n$. $I$ is the identity matrix. Then we can use the reparameterization trick\cite{VAE} to write $p_{\theta}(\bm{z}^n|\bm{X}^n)d\bm{z}^n=p(\bm{\epsilon})d\bm{\epsilon}$ and $\bm{z}^n=\bm{\mu}(\bm{X}^n)+\bm{\sigma}(\bm{X}^n)\cdot\bm{\epsilon}=\bm{\mathcal{E}}(\bm{X}^n,\bm{\epsilon};\theta)$, where $\bm{\epsilon}\sim\mathcal{N}(0,I)$ and the encoder function $\bm{\mathcal{E}}$ is a deterministic nonlinear function parameterized by a neural network.

\subsection{Dependence of SPIB on $\Delta t$}
\label{sec:time_dependence_SPIB}
In RAVE\cite{pRAVE} as well as in this algorithm, the time delay $\Delta t$ plays an important role in the simplification of the learning process. A time delay $\Delta t = 0$ is tantamount to ignoring the dynamics completely and simply clustering the input configuration into different states, while $\Delta t > 0$ can filter out all the fast modes, helping us ignore unnecessary details of the dynamical processes. Given its critical importance, in this section we analyze $\Delta t$ in detail.

Given a Markov process $\bm{X}_t$, if the initial probability distribution is given by $\rho_0$, the corresponding probability distribution after a lag time $\tau$ is 
\begin{equation} 
\label{eq:propagator}
\rho_{\tau}(\bm{X})=\int \rho_0(\bm{X}')P_{\tau}(\bm{X}|\bm{X}')d\bm{X}'\equiv \mathcal{P}(\tau) \rho_{0}
\end{equation}
where the time evolution of the probability density is governed by a linear operator $\mathcal{P}(\tau)$, called the propagator for the process $\bm{X}_t$. To explain the role of the time delay $\Delta t$ in our algorithm, we can derive a spectral decomposition for this operator $\mathcal{P}(\tau)$ by assuming the dynamics is reversible:\cite{Spectral_Decomp}
\begin{equation} 
\label{eq:spectral_decomp}
\rho_{\tau}=v_1+\sum_{i=2}^{\infty}a_i(\rho_0)\lambda_i(\tau)v_i 
\end{equation}
where $\{v_i\}$ are the propagator’s eigenfunctions and $\lambda_i=\exp{(-k_i\tau)}$ are the eigenvalues which decay exponentially in time with rates $k_i$. In principle, SPIB will ignore the dynamical processes whose timescale $t_i=1/k_i$ is comparable to or even smaller than the time delay $\Delta t$, as its corresponding component in Eq. \ref{eq:spectral_decomp} will decay exponentially. Thus, we interpret the time delay $\Delta t$ as the minimal time resolution that we seek to maintain for the dynamical system.

As discussed in the previous subsection, SPIB predicts from the present configuration $\bm{X}^n$ the future state $\bm{y}^{n+k}$ instead of the exact configuration $\bm{X}^{n+k}$. A subtle assumption made to justify this simplification is that the fluctuation inside each metastable state should be much faster than the transitions between different states. Such a timescale separation allows us to rewrite Eq. \ref{eq:spectral_decomp} as a sum of the stationary state eigenvector $v_1$ and two other parts:
\begin{equation} 
\begin{aligned} 
\label{eq:spectral_decomp2}
\rho_{\tau}&=v_1+\sum_{i=2}^{m}a_i(\rho_0)\lambda_i(\tau)v_i+\sum_{i=m+1}^{\infty}a_i(\rho_0)\lambda_i(\tau)v_i\\
&=v_1+\sum_{i=2}^{m}a_i(\rho_0)\lambda_i(\tau)v_i+\mathcal{P}_{fast}(\tau)\rho_0.
\end{aligned} 
\end{equation}
The first $m$ slow processes $\{v_i\}_{i=1}^m$ correspond to the state-to-state transitions that we are interested in, while the fast processes $\mathcal{P}_{fast}(\tau)$ represent the motions related to the molecular relaxation within these states. Therefore, an appropriate time delay $\Delta t$ should satisfy $t^{m+1}<\Delta t\ll t^{m}$ in order to screen out all the fast processes. In practice, this can be checked by examining the robustness of the results against different values of $\Delta t$, as we show in Sec. \ref{sec:Results}.

\subsection{Discrete-State Representation and Iterative Retraining Algorithm}
\label{sec:iterative_retraining_algorithm}

The SPIB framework introduced thus far requires a prior knowledge of states in the system, which is usually intractable especially for complex systems. To surmount this limitation, here we introduce an iterative technique to obtain a converged discrete-state representation based on the selected time delay $\Delta t$ we described in Sec. \ref{sec:time_dependence_SPIB}. The central idea is that if one configuration was located at state $i$ at a certain time, then after time delay $\Delta t$, it should still have the largest probability to be found at state $i$, since $\Delta t$ is much shorter than the typical escape time from a metastable state.

Given a set of initial state labels $\{\bm{y}^n\}$, we can write down the optimal predictor $\bm{\hat{y}}^*$ by assuming ergodic dynamics and setting it equal to a vector of probabilities $\bm{K}^* = \left\{ K_i^*(\bm{X};\Delta t) \right\}$:
\begin{equation} 
\begin{aligned} 
\label{eq:state_transition_density}
K_i^*(\bm{X};\Delta t)&=\frac{1}{\rho(\bm{X})}\lim_{T \to +\infty}\int_{0}^{T} h_i (\bm{X}_{t+\Delta t})\delta(\bm{X}-\bm{X}_t)dt\\
\text{where}\ &\rho(\bm{X})=\lim_{T \to +\infty}\int_{0}^{T}\delta(\bm{X}-\bm{X}_t)dt
\end{aligned} 
\end{equation}
Here $\bm{h}(\bm{X})=\{h_i(\bm{X}) \text{ for } i \in \left[ 1,D\right]\}$ is the state label function that maps the trajectory $\{\bm{X}^n\}$ to the $D$-dimensional state labels $\{\bm{y}^n\}$, and $\rho(\bm{X})$ represents the equilibrium density of $\bm{X}$. $K_i^*(\bm{X};\Delta t)$ can be  interpreted as the probability that the system starting from $\bm{X}$ will be found in state $i$ after a time delay $\Delta t$. As it is a function of the input configuration $\bm{X}$ and represents a state-transition probability, we call the function $\bm{K}^*(\bm{X};\Delta t)$ as the state-transition density. 

With this set-up, we now introduce a simple iterative scheme that is at the heart of our SPIB approach, as it allows us to learn the number and location of states on the fly with minimal human intervention. We start with an arbitrary set of labels $\{1,...,D\}$ for the system, where both the number and location of labels are some initial guess. If the system initiated from a certain high-dimensional configuration $\bm{X}$ has the largest probability to be found after time delay $\Delta t$ in some state $i$ from these initial labels, then  the label of the configuration $\bm{X}$ will be refined and updated to state $i$. We can denote the deterministic output of SPIB as $\bm{\hat{y}}=\bm{K}(\bm{X};\Delta t, \theta)\equiv \bm{\mathcal{D}}(\bm{\mu}(\bm{X});\Delta t, \theta)$, which tries to approximate the best predictor $\bm{\hat{y}}^*=\bm{K}^*(\bm{X};\Delta t)$. Then a set of new state labels can be generated by:
\begin{equation} 
\label{eq:label_update}
h_i(\bm{X})=\left\{\begin{aligned} 
&1\ (i=\underset{j}{\text{argmax}}\ K_j(\bm{X};\Delta t,\theta))\\
&0\ (\text{otherwise})\\
\end{aligned} \right.\ \text{for}\ i=1,\ldots,D.
\end{equation}
This label refinement step might very well lead to null assignments for some of the labels we started with, as shown in Sec. \ref{sec:Results} for actual test cases. 

\begin{figure}[ht]
    \centering
    \includegraphics[width=0.45\textwidth]{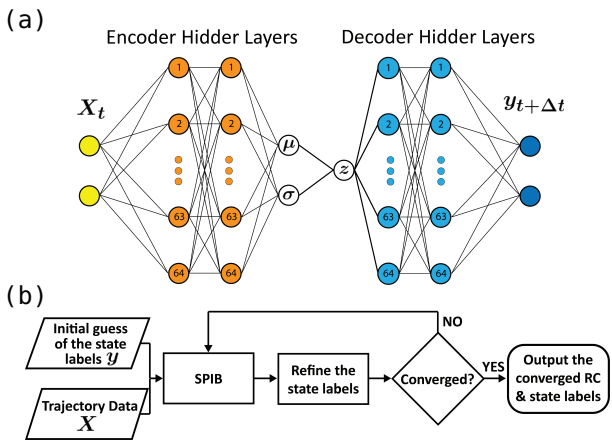}
    \caption{ (a) Network architecture used for SPIB. Both the encoder and decoder are nonlinear deep neural networks. (b) A flowchart illustrating SPIB.}
    \label{fig:alg}
\end{figure}

Based on Eq. \ref{eq:state_transition_density} and \ref{eq:label_update}, an iterative retraining can be performed and the whole algorithm is summarized through Fig. \ref{fig:alg}(b) and Alg. \ref{algo:SPIB}. Thus, as illustrated above, we expect such a converged discrete-state representation $\bm{h}(\bm{X})$, by this self-consistent design, should only depend on the dynamic properties of the system and the time delay $\Delta t$. Moreover, on account of the screening property of the time delay $\Delta t$, the final representation will automatically ignore transient intermediate states and only figure out those long-lived metastable states. This in fact offers us a powerful tool to obtain a dynamics-based coarse-grained description of the complex system.

\begin{algorithm}[H]
    \captionsetup{justification=centering}
    \caption{SPIB}\label{algo:SPIB}
    \begin{algorithmic}[1]
        \Require a long unbiased trajectory $\{\bm{X}^n\}$, a set of initial state labels $\{\bm{y}^n\}$, RC dimensionality $d$, the number of pseudo-inputs $K$, time delay $\Delta t$ 
        \Repeat
            \For{$i$ in $m$}
                \State{Sample a minibatch $\{\bm{X}^n\}$ and $\{\bm{y}^n\}$}
                \State{Calculate the objective function $\mathcal{L}$}
                \State{Update the neural network parameters $\theta$, pseudo-inputs $\{\bm{u}^k\}_{k=1}^{K}$, pseudo-weights $\{\omega_k\}_{k=1}^{K}$} 
            \EndFor
            \State{Update the state labels $\{\bm{y}^n\}$ by Eq. \ref{eq:label_update}}
        \Until{convergence of RC, state-transition density, and state labels}
    \end{algorithmic}
\end{algorithm}

\subsection{State-Transition Density and Committor}
\label{sec:state_transition_density}

Recently a few methods have been proposed to calculate the committor through the construction of Markov state models (MSM).\cite{Markov_model,transition_networks,Wales2009,Lane2011,Galerkin_approximation} By constructing an efficient MSM, the committor can be calculated directly from the transition matrix by solving a system of linear equations.\cite{transition_networks} However, this usually requires a large number of discrete states in order to estimate the MSM transition matrix and thus the committor accurately \cite{Wales2009,Lane2011,Galerkin_approximation}, thereby severely diminishing the interpretability of the model. This also means requiring a very well sampled trajectory moving accurately capturing transitions between the large number of different states, which might be hard to achieve.\cite{biswas2018metadynamics}  As we will show in this section, our SPIB approach can efficiently estimate the transition density by relegating the need to know the exact transition probabilities within the metastable states of the system. In other words, we will demonstrate that SPIB can learn the approximate committor and identify correct transition state regions even with a small number of discrete states relative to MSM type approaches. 

As discussed in the Sec. \ref{sec:time_dependence_SPIB}, the timescale separation between the state-to-state transitions and the fluctuations inside each metastable state allows the factorization of the transition density $P_{\Delta t}(\bm{X}|\bm{X}_0)$ into two parts: 
\begin{equation} 
\begin{aligned} 
\label{eq:assumption}
P_{\Delta t}(\bm{X}|\bm{X}_0)=\sum_{i=1}^{D} P_i(\bm{X}) K_i^*(\bm{X}_0;\Delta t)\\
\text{where}\ P_i(\bm{X})= \frac{\rho(\bm{X})h_i(\bm{X})}{\int h_i(\bm{X}) \rho(\bm{X}) d\bm{X}}.
\end{aligned} 
\end{equation}
In Eq. \ref{eq:assumption}, the first part $P_i(\bm{X})$ represents the equilibrium probability density for any state $i \in \left[ 1,D\right]$, while the second part $K_i^*(\bm{X}_0;\Delta t)$ is exactly the state-transition density defined in Eq. \ref{eq:state_transition_density}. Given the definition of two states A and B whose committor attracts our interest, the committor $p_B(\bm{X})$ then can be obtained by solving following linear integral equations:
\begin{equation} 
\begin{aligned} 
\label{eq:committor_state_transition_density}
p_B(\bm{X})&=\int p_B(\bm{X}')P_{\Delta t}(\bm{X}'|\bm{X})d\bm{X}' &if\ \bm{X}\notin A\cup B\\
p_B(\bm{X})&=1 &if\ \bm{X}\in B\\
p_B(\bm{X})&=0 &if\ \bm{X}\in A.\\
\end{aligned} 
\end{equation}
However, instead of solving Eq. \ref{eq:committor_state_transition_density} explicitly and tabulating configurations where $p_B(\bm{X}) \approx 0.5 $, the transition state ensemble (TSE) can also be identified using this state-transition density directly. If a set of trajectories starting from some $\bm{X}$ have the largest, approximately equal probabilities of transitioning to two different states after time delay $\Delta t$, then $\bm{X}$ can be considered to belong to the TSE. Through the numerical examples in the section \ref{sec:Results}, we will illustrate that this new definition of transition states is, in fact, as valid as the original committor-based definition. Besides, we would also like to highlight that based on this definition and Alg. \ref{algo:SPIB}, the ensemble of transition states will eventually form the boundaries of the finally converged discrete states. Therefore, the state-transition density $\bm{K}^*(\bm{X};\Delta t)$ provides us with an accurate but intuitive way to characterize the transitions between different metastable states. 

Overall, we believe that such a state-transition density finally generated by our algorithm is a reasonable substitute for the committor as it can quantitatively describe the dynamical behaviors of every states along a trajectory and further identify the correct transition states. Thus, the most informative representation $\bm{z}$ given by the encoder $p_{\theta}(\bm{z}^n|\bm{X}^n)$ about this state-transition density learned by SPIB should naturally serve as a reasonable RC approximating the committor. But for simplicity, in the following discussion, our RC will refer specifically to the deterministic part or the mean value $\bm{\mu}$ of the representation $\bm{z}$ to better compare with traditional deterministic prescriptions.

\section{Results}
\label{sec:Results}

\subsection{Model Systems}
\label{sec:model_systems}

\begin{figure}[ht]
    \centering
    \includegraphics[width=0.45\textwidth]{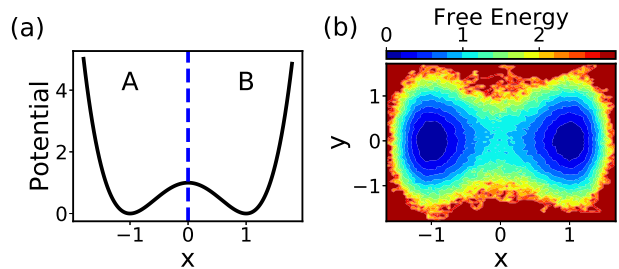}
    \caption{ Double-well analytical potentials projected along x-axis (a) and its corresponding probability $P(x,y)$ distribution of the generated trajectory (b), plotted as the free energy $-k_B T \text{ log }P(x,y)$.}
    \label{fig:DW_model_potential}
\end{figure}

To demonstrate our SPIB approach in practice, here we benchmark it for different model potentials, including two analytical potentials, and the small biomolecule alanine dipeptide in vacuum. The first analytical potential $U_{DW}(x,y)$ comprises a double well in two dimensions, shown in Fig. \ref{fig:DW_model_potential}. The second potential is made of four wells also in two dimensions, shown in Fig. \ref{fig:FW_model_potential}. The governing potentials are given by
\begin{equation} 
\label{eq:DW}
U_{DW}(x,y)=(x^2-1)^2+y^2
\end{equation}
and
\begin{equation} 
\label{eq:FW}
\begin{aligned}
U_{FW}(x,y)=&2[x^8+0.6e^{-80x^2}+0.2e^{-80(x-0.5)^2}\\
&+0.5e^{-40(x+0.5)^2}]+(x^2-1)^2+y^2.
\end{aligned}
\end{equation}

The trajectories for these two potentials were generated using Langevin dynamics simulation\cite{Langevin} with integration timestep of 0.001 units, inverse temperature ${(k_B T)}^{-1} =3.0$ and friction coefficient $\gamma=4.0$, where $k_B$ is Boltzmann constant. For either potential, we used a long equilibrium trajectory equaling 60,000 time units with a temporal resolution of 0.01 units. 

For the study of conformation transitions in alanine dipeptide in vaccum, the simulation was performed with the software GROMACS 5.0,\cite{GROMACS1995,GROMACS2015} patched with PLUMED 2.4.\cite{PLUMED2} The temperature was kept constant at 450 K using the velocity rescaling thermostat\cite{v_rescaling} and the integration time step was 2 fs. A 800 ns long equilibrium trajectory with a temporal resolution of 0.01 ps was employed to train and test our algorithm.

\subsection{Neural network architecture and training}
In this paper, both the encoder and decoder are nonlinear and parameterized by fully connected neural networks with two hidden layers as shown in Fig. \ref{fig:alg}(a). Each hidden layer in both the encoder and decoder has 16 nodes for the two analytical potentials, and 64 nodes for alanine dipeptide. All these hidden layers use a rectified linear unit (ReLU) as the activation function.

The networks were trained using the Adam optimizer\cite{Adam} with a learning rate of $0.001$ and a batch size of 2048 for all the numerical examples. The state labels are refined every 1000 training steps for analytical potentials, and every 2000 training steps for alanine dipepetide.

\subsection{Double-Well Analytical Potential}
\label{sec:double_well}

We first demonstrate SPIB for the double-well analytical potential. For this first example, we assume we already know the system relatively well by setting the RC dimension $d=1$, the number of pseudo-inputs $K=2$ and the dimension of state labels $D=2$. In following sections \ref{sec:four_well} and beyond, we remove the need for making any such assumptions and show how SPIB still works very well. For this double well system, in order to generate an initial guess of state labels, the samples are labeled as state A if $x<b$ and B otherwise. Here, the initial boundary point $b$ can be changed to test the robustness of SPIB. 

\begin{figure}[ht]
    \centering
    \includegraphics[width=0.45\textwidth]{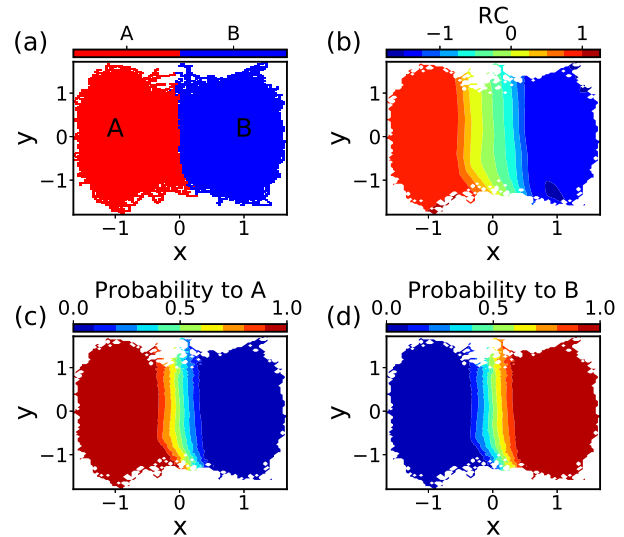}
    \caption{The results of SPIB for double-well potential. (a) The converged state labels A and B. (b) Different values of the RC illustrated in the $x-y$ plane. (c) and (d) are the state-transition density learned by SPIB, where (c) represents the transition density to state A and (d) represents the transition density to state B.}
    \label{fig:DW_results}
\end{figure}

The final converged results are shown in Fig. \ref{fig:DW_results}. Fig. \ref{fig:DW_results}(a) illustrates that SPIB can learn the correct state labels, where the boundary is located at around $x=0$. In Fig. \ref{fig:DW_results}(b), as the $y$-direction is pure noise, the learned RC is almost independent of the $y-$direction, suggesting that SPIB is able to distinguish important features from noise. Besides, Fig. \ref{fig:DW_results}(b) also shows that as desired, the fluctuations inside each state are not captured by SPIB, as they are almost mapped to a single point in RC. Fig. \ref{fig:DW_results}(c) and (d) present the state-transition density learned by our algorithm, which is highly correlated with our RC. 

\begin{figure}[ht]
    \centering
    \includegraphics[width=0.45\textwidth]{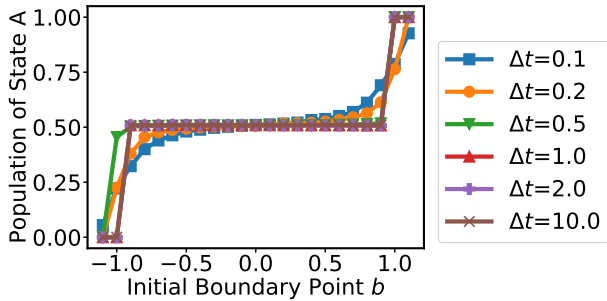}
    \caption{The robustness of SPIB in double-well analytical potential. The x-axis represents the initial boundary point $b$ while y-axis represents the converged fractional population of state A ($\sum_{j=1}^{N}y^j_A/N$). The lines start to overlap especially when $\Delta t  \gtrsim 0.5$.} 
        \label{fig:DW_robustness}
\end{figure}

We now further demonstrate that our results obtained above are robust to the initial boundary demarcating parameter $b$ and the time delay $\Delta t$. As shown in Fig. \ref{fig:DW_robustness}, a large range of  $b$ and $\Delta t$ values can result in the same state definition. The fractional population of state A is defined by the ratio of the number of samples finally labeled as state A to the total number of samples ($f_A=\sum_{j=1}^{N}y^j_A/N$). For the initial boundary point $b$, the only constraint is that it should not be large than 1 or smaller than -1; otherwise, state A and state B will be regarded as one state by our algorithm. $\Delta t$ can be anywhere between the molecular relaxation time scale ($\Delta t  \gtrsim 0.5$) and the interconversion timescale between state A and B, which is around the implied timescale of $t_1=54$ shown in Supplementary Material (SM). 

Here, all the results are obtained by setting  the hyper-parameter $\beta=0.03$ in Eq. \ref{eq:SPIB_obj} (not to be confused with the inverse temperature ${(k_B T)}^{-1}$ ), which can be determined by choosing the turning point on the Rate-Distortion plot (see SM). However, as long as $\beta$ is not too large, we found our results are still very robust to the selection of $\beta$. 

\subsection{Four-Well Analytical Potential}
\label{sec:four_well}

\begin{figure}[ht]
    \centering
    \includegraphics[width=0.46\textwidth]{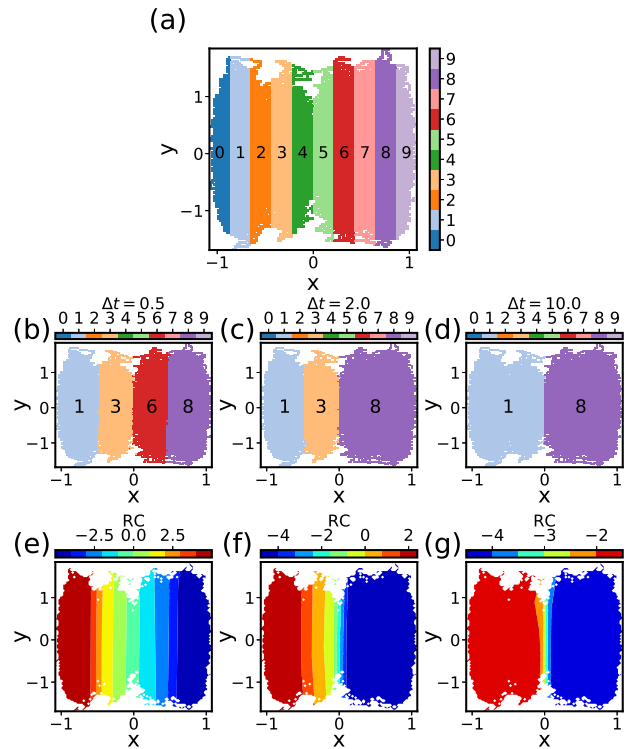}
    \caption{The time delay dependent discrete-state representation of the four-well potential model. The initial state labels are shown in (a), while the converged results for different time delays are presented in (b-g). Middle row shows the state labels for different time delays while bottom row shows the corresponding RC. The state labels and RC were learned using the time delay 0.5 (b,e), 2 (c,f), and 10 (d,g) respectively.}
    \label{fig:FW_results}
\end{figure}

We now apply SPIB to a four-well analytical potential where we do not assume any prior knowledge about the system such as the number of metastable states. In this case, we arbitrarily discretized the input data space into sufficiently fine grids as our initial state labels, shown in Fig. \ref{fig:FW_results}(a). We set the RC dimension $d=1$, the number of pseudo-inputs $K=10$ and the dimension of state labels $D=10$. In other words, here we have deliberately taken $K$ and $D$ to be arbitrarily large relative to the true number of metastable states. In order to let the RC only contain the important information, we chose $\beta=0.01$ (see SM).

\begin{figure}[ht]
    \centering
    \includegraphics[width=0.45\textwidth]{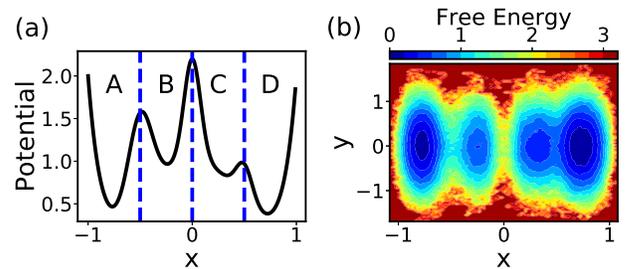}
    \caption{Four-well analytical potentials projected along x-axis (a) and its corresponding probability $P(x,y)$ distribution of the generated trajectory (b), plotted as the free energy $-k_B T \text{ log }P(x,y)$.}
    \label{fig:FW_model_potential}
\end{figure}

\begin{figure}[ht]
    \centering
    \includegraphics[width=0.45\textwidth]{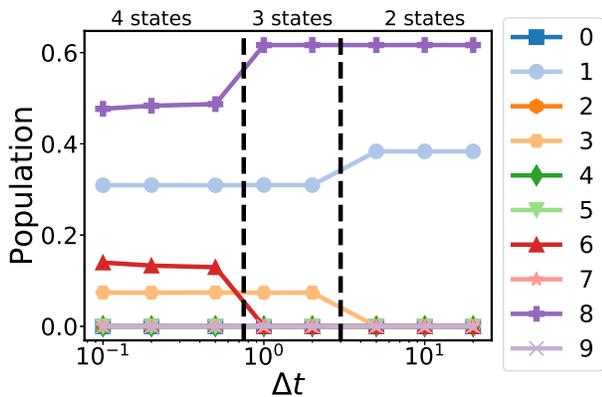}
    \caption{The robustness of SPIB on four-well analytical potential can be seen by plotting the fractional population of different states ($f_i=\sum_{j=1}^{N}y^j_i/N\ \ for$ $i=0,\cdots,9$). With different time resolutions (or time delays $\Delta t$), the system is coarse grained into four states, three states and two states.}
        \label{fig:FW_robustness}
\end{figure}

\begin{figure}[ht]
    \centering
    \includegraphics[width=0.46\textwidth]{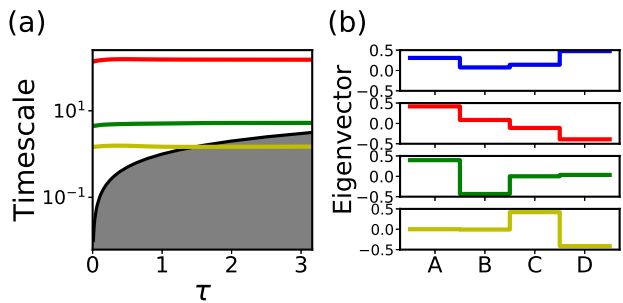}
    \caption{The implied timescales (a) and corresponding eigenvectors (b) for the four-well analytical potential. (a) The converged values of the implied timescales are $t_2=151.7$ (red), $t_3=5.3$ (green), $t_4=1.5$ (yellow). The grey area under the black line represents the timescale that is smaller than the lag time $\tau$. (b) The first eigenvector (blue line) represents the stationary probability distribution; the second eigenvector (red line) mainly represents the transition between state A and state D; the third eigenvector (green line) represents the transition between state A and state B; the last eigenvector (yellow line) represents the transition between state C and state D.}
    \label{fig:FW_implied_timescale}
\end{figure}

Fig. \ref{fig:FW_results} shows the state labels and RC learned by SPIB using different time delays $\Delta t$. There are several interesting observations that can be made here. Firstly, in Fig. \ref{fig:FW_results}(b), we find that SPIB can still obtain the correct state labels by choosing an appropriate time delay ($\Delta t=0.5$) without any prior information. This is very promising for practical problems as a precise state definition or even the number of states are usually unavailable in complex systems. Secondly, we also find that a dynamically truthful discrete-state representation can be obtained by SPIB using different time delays $\Delta t$. When the time delay increases ($\Delta t=2.0$), the original state
C and state D shown in Fig. \ref{fig:FW_model_potential}(a) cannot be distinguished by SPIB any more (Fig. \ref{fig:FW_results}(c)). If we further increase the time delay ($\Delta t=10$), even the original state A and state B will become indistinguishable (Fig. \ref{fig:FW_results}(d)). The time dependence of state labels can be explained by the different timescales of transitions between states (Fig. \ref{fig:FW_implied_timescale}). These results unequivocally shed light into the role of the time delay $\Delta t$ in our algorithm--it filters out the fast modes of dynamics, and provides a dynamics-based coarse-gained understanding of the complex system. We also point out that although our results depend on the selection of time delay, they are in fact still very robust to changes of $\Delta t$. Fig. \ref{fig:FW_robustness} shows a broad range of $\Delta t$ can result in the same discrete-state representation.

\subsection{Alanine Dipeptide in Vacuum}
\label{sec:aladip}

\begin{figure}[ht]
    \centering
    \includegraphics[width=0.45\textwidth]{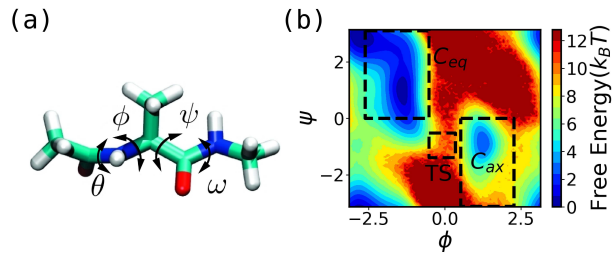}
    \caption{(a) Alanine dipeptide molecule illustrated along with four dihedral angles: $\phi$ ($C$-$N$-$C_\alpha$-$C$), $\psi$ ($N$-$C_\alpha$-$C$-$N$), $\theta$ ($O$-$C$-$N$-$C_\alpha$), $\omega$ ($C_\alpha$-$C$-$N$-$C$). (b) The generated free energy surface of alanine dipeptide in vaccum at 450K along the dihedral angles $\phi$ and $\psi$.  The regions described in boxes are usually defined as state $C_{eq}$: ($-150^{\circ}\le\phi\le -30^{\circ}$, $0^{\circ}\le\psi\le 180^{\circ}$), $C_{ax}$: ($30^{\circ}\le\phi\le 130^{\circ}$, $-180^{\circ}\le\psi\le 0^{\circ}$) and approximate TS: ($-30^{\circ}\le\phi\le 20^{\circ}$, $-80^{\circ}\le\psi\le -30^{\circ}$).\cite{committor_aladip}}
    \label{fig:aladip}
\end{figure}

\begin{figure}[ht]
    \centering
    \includegraphics[width=0.46\textwidth]{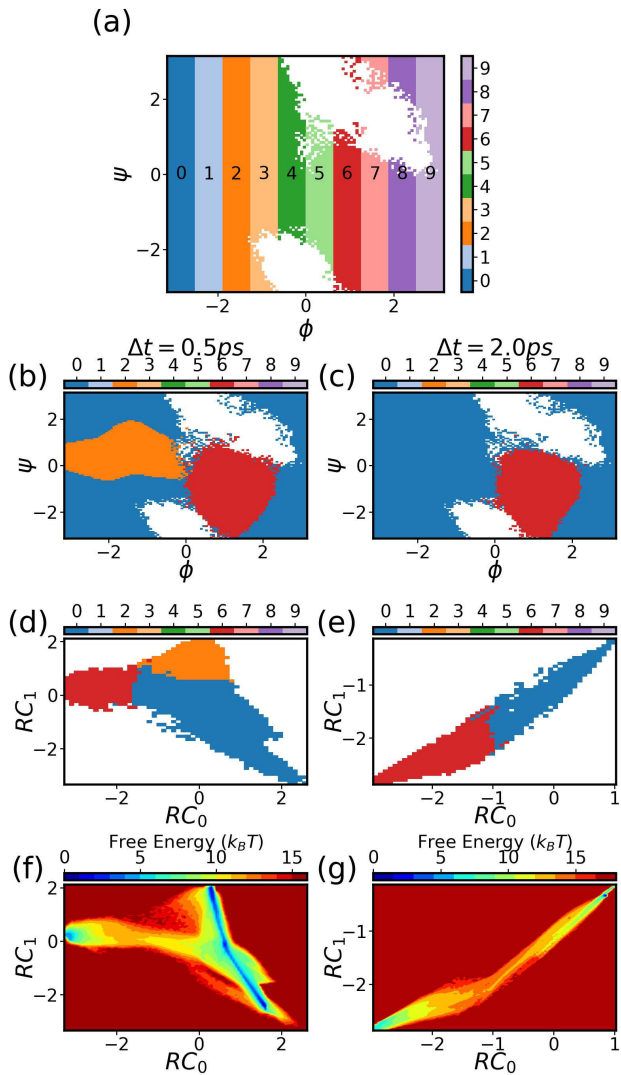}
    \caption{The time-dependent discrete-state representation of alanine dipeptide in vacuum. The initial state labels are shown in (a). A three-state representation was learned by using the time delay $\Delta t=0.5ps$ (b,d,f), and a two-state representation was obtained by using the time delay $\Delta t=2ps$ (c,e,f). The second row (b,c) are the state labels projected to $\phi$-$\psi$ space. The color (or state label) in each grid corresponds only to the state label with highest fraction of samples for the respective grid point. The third row (d,e) are the state labels learned in the 2D RC space. The fourth row (f,g) shows the free energy surface (-$k_BT\log P(RC_1,RC_2)$) in the 2D RC space.}
    \label{fig:aladip_result}
\end{figure}

Finally, we employ SPIB to study conformation transitions in the small biomolecule alanine dipeptide. The trajectory here was expressed in terms of four dihedral angles $\phi$, $\psi$, $\theta$ and $\omega$, illustrated  in Fig. \ref{fig:aladip}. Here we discretized the input data space along $\phi$ into 10 grids as our initial state labels, as shown in Fig. \ref{fig:aladip_result}(a), and set the RC dimension $d=2$, the number of pseudo-inputs $K=10$ and the dimensionality of state labels $D=10$. $\beta=0.01$ was chosen to generate the most informative RC. 

\begin{figure}[ht]
    \centering
    \includegraphics[width=0.45\textwidth]{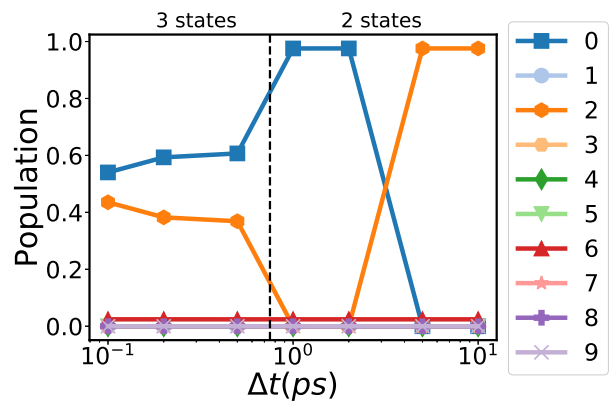}
    \caption{The robustness of SPIB on alanine dipeptide through the fractional population of different states ($f_i=\sum_{j=1}^{N}y^j_i/N\ \ for\ i=0,\cdots,9$). With different time resolutions (or time delays $\Delta t$), the system is coarse grained into three states (e.g. $f_0=0.61,$ $f_2=0.37,$ $f_6=0.02$ at $\Delta t=0.5\ ps$) and two states (e.g. $f_0=0.98,f_6=0.02$ at $\Delta t=2\ ps$). Though it might appear that there is a flip between the state label $0$ and $2$ when $\Delta t>2\ ps$, the same metastable states are obtained.}
    \label{fig:aladip_robustness}
\end{figure}

\begin{figure}[ht]
    \centering
    \includegraphics[width=0.45\textwidth]{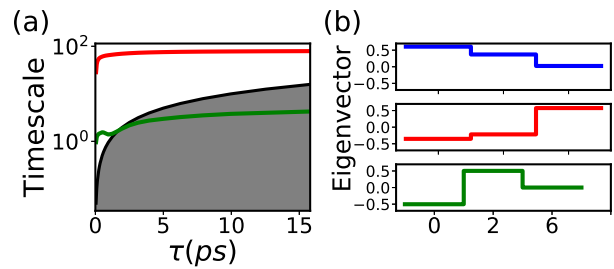}
    \caption{The implied timescales (a) and corresponding eigenvectors (b) of alanine dipeptide.(a) The converged implied timescales $t_2=79ps$ (red), $t_3=4.2ps$ (green). The grey area under the black line represents the timescale that is smaller than the lag time $\tau$. (b) The first eigenvector (blue line) represents the stationary probability distribution; the second eigenvector (red line) mainly represents the transition between state $C_{ax}$ (state 6) and state $C_{eq}$ (state 0/2); the third eigenvector (green line) represents the transition inside state $C_{eq}$ (or between state 0 and state 2).}
    \label{fig:aladip_implied_timescale}
\end{figure}

Similar to our previous results for four-well analytical potential, in Fig. \ref{fig:aladip_result}(b), we show how SPIB can still learn successfully the state labels corresponding to the three well-known free energy minima in the $\phi$-$\psi$ space shown in Fig. \ref{fig:aladip}(b). When the time delay $\Delta t$ = $2ps$, the two free energy minima located in the top-left corner of the $\phi$-$\psi$ space (Fig. \ref{fig:aladip}(b)) become indistinguishable from a dynamical perspective, given that the interconversion times between these two metastable states is now close to $\Delta t$ (Fig. \ref{fig:aladip_implied_timescale}). Thus, only 2 states are obtained in Fig. \ref{fig:aladip_result}(c). We then further demonstrate in Fig. \ref{fig:aladip_robustness} that such a coarse-grained understanding obtained by SPIB is still very robust in alanine dipeptide, as the same state labels are obtained with a broad range of $\Delta t$. 

\begin{figure}[ht]
    \centering
    \includegraphics[width=0.45\textwidth]{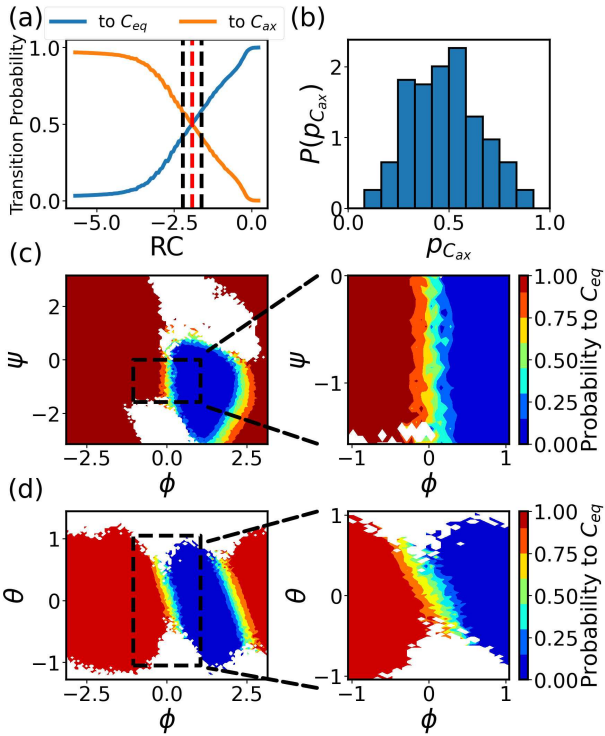}
    \caption{(a) The 1D RC of alanine dipeptide learned by SPIB using $\Delta t=2ps$. The red dotted line represents the transition state ($RC^*=-1.92$), while the black dotted lines shows the neighborhood in vicinity of the transition state range $[-2.22<RC<-1.62]$. In (b) we show the committor probability distribution of $p_{C_{ax}}$ under the constraint $-2.22<RC<-1.62$. (c) and (d) represent the state-transition density to $C_{ax}$ projected to $\phi$-$\psi$ space and $\phi$-$\theta$ space respectively. {In the bottom two rows,} the figures in the right column are the zoomed plots around the TS region defined by Fig. \ref{fig:aladip}(b).}
    \label{fig:aladip_d=1}
\end{figure}

The 2-D RC so learnt through SPIB is presented in Fig. \ref{fig:aladip_result}(d-g). The free energy surface in the RC space shown in Fig. \ref{fig:aladip_result}(f,g) indicates that the barrier between $C_{eq}$ and $C_{ax}$ defined in Fig. \ref{fig:aladip}(b) is much higher than the barrier between the two local minima (state $0$ and state $2$ in Fig. \ref{fig:aladip_result}(b,d)) within $C_{eq}$. This also explains why different interconversion timescales are obtained in Fig. \ref{fig:aladip_implied_timescale}. From these results, we can see that a 1-D RC is enough if we just want to identify the transitions between $C_{eq}$ and $C_{ax}$. Thus, we reran the analysis of $\Delta t=2ps$ with the RC dimension $d=1$ and show the new RC so-obtained in Fig. \ref{fig:aladip_d=1}(a). Such a 1D RC can be easily used to identify the transition state, which has the same state transition probability to two different metastable states. In Fig. \ref{fig:aladip_d=1}(a), the transition state corresponds to $RC=-1.92$. To test whether our RC can identify the correct transition states, we chose to focus on the states located in the TS region shown in Fig. \ref{fig:aladip}(b). By doing a traditional, detailed committor analysis, we obtained the reference committor. For this we launched 50 1-ps trajectories with random initial Maxwell-Boltzmann velocities for each configuration in vicinity of the TS under the constraint $-2.22<RC<-1.62$, and then calculated their committor function $P_{C_{ax}}$ based on the fraction of trajectories reaching $C_{ax}$ prior to $C_{eq}$. This committor probability distribution is shown in Fig. \ref{fig:aladip_d=1}(b), where it can be seen clearly that the probability of $p_{C_{ax}}$ is characterized by a single peak centered at $p_{C_{ax}}\sim0.5$. This shows that the RC from SPIB indeed meets the traditional expectations from a RC.\cite{best2005reaction}

We now show in Fig. \ref{fig:aladip_d=1}(c,d) our learned state-transition density to the $C_{ax}$ state projected in the $\phi$-$\psi$ plane and $\phi$-$\theta$ plane. Fig. \ref{fig:aladip_d=1}(c) shows that the transition states in TS region are aligned almost parallel to the $\psi$ axis, suggesting they are in fact almost irrelevant to $\psi$. Fig. \ref{fig:aladip_d=1}(d), however, indicates that both $\phi$ and $\theta$ are required in order to identify the transition state. Both of these findings are in good agreement with previous reports for this system.\cite{committor_aladip,Ma2005,Mori2020} Thus, our results confirm that the state-transition density generated by SPIB can be a reasonable substitute for the committor, and the RC learned can capture the most important dynamical information from the input trajectory to identify the correct transition states.

\section{Discussion}
In this work, we have proposed a deep learning based algorithm called State Predictive Information Bottleneck (SPIB) to learn the RC from trajectory data. SPIB builds up on the insights we have introduced previously in the RAVE family of methods,\cite{RAVE,pRAVE} and by changing the nature of the information bottleneck based objective function, it allows generating new physical insights from typically hard to interpret deep neural networks. We have first showed that the time delay $\Delta t$ can be interpreted as the time resolution that we care about in a dynamical system, and through this we can control the degree of coarse-graining obtained by our algorithm. Once a time delay $\Delta t$ is selected, SPIB can automatically index the high-dimensional state space into metastable states through an iterative retraining algorithm, and then characterize their dynamic behaviors in terms of state-transition density. This provides us with a promising way to analyze generic complex systems and interpret the massive data generated by MD simulations.

We have also demonstrated that the bottleneck variable learned in SPIB tries to carry the maximum information of the state-transition density, which in principle can be equivalent to the traditional committor function, if there is a timescale separation between the state-to-state transitions and the fluctuations within metastable states. Then through numerical tests on benchmark systems, we confirmed that the state-transition density generated by SPIB is a reasonable substitute for committor, and demonstrated that our RC can focus only on the motion related to state transitions and capture the most important features from trajectories to identify the correct transition states. 

We finish this section by describing some of the exciting new avenues pertaining to SPIB that we will explore in future work. First pertains to the use of SPIB for enhanced sampling. By choosing an appropriate time delay, the RC learned by SPIB can correctly identify different metastable states and even the transition states among them, which can be crucial to obtain better sampling.\cite{Bussi2020} This would involve reweighting input biased trajectories, obtained by biasing along some trial RC for instance. The reweighting can then be performed in the manner described in Ref. \onlinecite{RAVE,pRAVE}. Secondly, it is natural to desire that the RC learned by SPIB can be interpreted in terms of a few human-understandable physical variables. However, unlike RAVE\cite{pRAVE} here we use a nonlinear deep neural network as our encoder, and thus the interpretation of our RC is not a trivial task. In order to deal with this question of interpretability, we will make use of approaches in representation learning.\cite{Bengio2013} Overall, we believe our algorithm is a step towards a more complete understanding of complex systems by making use of the variability offered by the information bottleneck framework of AI,\cite{IB,variational_IB,alemi2018therml} and should be useful to a broad range of scientific communities. \newline

\textbf{Supplementary material\newline }
See supplementary material for other numerical details. \newline

\textbf{Acknowledgements\newline }
This research was entirely supported by the U.S. Department of Energy, Office of Science, Basic Energy Sciences, CPIMS Program, under Award DE-SC0021009. The authors thank Sun-Ting Tsai for sharing the code implementing Langevin dynamics, Luke Evans for sharing the GROMACS script, Yihang Wang and Zachary Smith for in-depth discussions. The authors also thank MARCC and XSEDE for providing computational resources used in this work.\newline

\textbf{Code availability statement\newline }
The python code of SPIB using Pytorch will be made available for public use at https://github.com/tiwarylab/State-Predictive-Information-Bottleneck. \newline

\textbf{Data availability statement\newline }
The data that support the findings of this study are available from the corresponding author upon reasonable request. \newline

\textbf{References}
\bibliography{references}

\begin{thebibliography}{46}%
\makeatletter
\providecommand \@ifxundefined [1]{%
 \@ifx{#1\undefined}
}%
\providecommand \@ifnum [1]{%
 \ifnum #1\expandafter \@firstoftwo
 \else \expandafter \@secondoftwo
 \fi
}%
\providecommand \@ifx [1]{%
 \ifx #1\expandafter \@firstoftwo
 \else \expandafter \@secondoftwo
 \fi
}%
\providecommand \natexlab [1]{#1}%
\providecommand \enquote  [1]{``#1''}%
\providecommand \bibnamefont  [1]{#1}%
\providecommand \bibfnamefont [1]{#1}%
\providecommand \citenamefont [1]{#1}%
\providecommand \href@noop [0]{\@secondoftwo}%
\providecommand \href [0]{\begingroup \@sanitize@url \@href}%
\providecommand \@href[1]{\@@startlink{#1}\@@href}%
\providecommand \@@href[1]{\endgroup#1\@@endlink}%
\providecommand \@sanitize@url [0]{\catcode `\\12\catcode `\$12\catcode
  `\&12\catcode `\#12\catcode `\^12\catcode `\_12\catcode `\%12\relax}%
\providecommand \@@startlink[1]{}%
\providecommand \@@endlink[0]{}%
\providecommand \url  [0]{\begingroup\@sanitize@url \@url }%
\providecommand \@url [1]{\endgroup\@href {#1}{\urlprefix }}%
\providecommand \urlprefix  [0]{URL }%
\providecommand \Eprint [0]{\href }%
\providecommand \doibase [0]{http://dx.doi.org/}%
\providecommand \selectlanguage [0]{\@gobble}%
\providecommand \bibinfo  [0]{\@secondoftwo}%
\providecommand \bibfield  [0]{\@secondoftwo}%
\providecommand \translation [1]{[#1]}%
\providecommand \BibitemOpen [0]{}%
\providecommand \bibitemStop [0]{}%
\providecommand \bibitemNoStop [0]{.\EOS\space}%
\providecommand \EOS [0]{\spacefactor3000\relax}%
\providecommand \BibitemShut  [1]{\csname bibitem#1\endcsname}%
\let\auto@bib@innerbib\@empty
\bibitem [{\citenamefont {Wang}, \citenamefont {Lamim~Ribeiro},\ and\
  \citenamefont {Tiwary}(2020)}]{ML_review}%
  \BibitemOpen
  \bibfield  {author} {\bibinfo {author} {\bibfnamefont {Y.}~\bibnamefont
  {Wang}}, \bibinfo {author} {\bibfnamefont {J.~M.}\ \bibnamefont
  {Lamim~Ribeiro}}, \ and\ \bibinfo {author} {\bibfnamefont {P.}~\bibnamefont
  {Tiwary}},\ }\href {\doibase 10.1016/j.sbi.2019.12.016} {\bibfield  {journal}
  {\bibinfo  {journal} {Curr Opin Struct Biol}\ }\textbf {\bibinfo {volume}
  {61}},\ \bibinfo {pages} {139} (\bibinfo {year} {2020})}\BibitemShut
  {NoStop}%
\bibitem [{\citenamefont {Bolhuis}, \citenamefont {Dellago},\ and\
  \citenamefont {Chandler}(2000)}]{committor_aladip}%
  \BibitemOpen
  \bibfield  {author} {\bibinfo {author} {\bibfnamefont {P.~G.}\ \bibnamefont
  {Bolhuis}}, \bibinfo {author} {\bibfnamefont {C.}~\bibnamefont {Dellago}}, \
  and\ \bibinfo {author} {\bibfnamefont {D.}~\bibnamefont {Chandler}},\ }\href
  {\doibase 10.1073/pnas.100127697} {\bibfield  {journal} {\bibinfo  {journal}
  {Proc Natl Acad Sci U S A}\ }\textbf {\bibinfo {volume} {97}},\ \bibinfo
  {pages} {5877} (\bibinfo {year} {2000})}\BibitemShut {NoStop}%
\bibitem [{\citenamefont {Bolhuis}\ \emph {et~al.}(2002)\citenamefont
  {Bolhuis}, \citenamefont {Chandler}, \citenamefont {Dellago},\ and\
  \citenamefont {Geissler}}]{committor_review}%
  \BibitemOpen
  \bibfield  {author} {\bibinfo {author} {\bibfnamefont {P.~G.}\ \bibnamefont
  {Bolhuis}}, \bibinfo {author} {\bibfnamefont {D.}~\bibnamefont {Chandler}},
  \bibinfo {author} {\bibfnamefont {C.}~\bibnamefont {Dellago}}, \ and\
  \bibinfo {author} {\bibfnamefont {P.~L.}\ \bibnamefont {Geissler}},\
  }\href@noop {} {\bibfield  {journal} {\bibinfo  {journal} {Advances in
  chemical physics}\ }\textbf {\bibinfo {volume} {53}},\ \bibinfo {pages} {291}
  (\bibinfo {year} {2002})}\BibitemShut {NoStop}%
\bibitem [{\citenamefont {Geissler}, \citenamefont {Dellago},\ and\
  \citenamefont {Chandler}(1999)}]{committor_ion}%
  \BibitemOpen
  \bibfield  {author} {\bibinfo {author} {\bibfnamefont {P.~L.}\ \bibnamefont
  {Geissler}}, \bibinfo {author} {\bibfnamefont {C.}~\bibnamefont {Dellago}}, \
  and\ \bibinfo {author} {\bibfnamefont {D.}~\bibnamefont {Chandler}},\
  }\href@noop {} {\bibfield  {journal} {\bibinfo  {journal} {The Journal of
  Physical Chemistry B}\ }\textbf {\bibinfo {volume} {103}},\ \bibinfo {pages}
  {3706} (\bibinfo {year} {1999})}\BibitemShut {NoStop}%
\bibitem [{\citenamefont {Pluharova}\ \emph {et~al.}(2016)\citenamefont
  {Pluharova}, \citenamefont {Baer}, \citenamefont {Schenter}, \citenamefont
  {Jungwirth},\ and\ \citenamefont {Mundy}}]{Pluharova2016}%
  \BibitemOpen
  \bibfield  {author} {\bibinfo {author} {\bibfnamefont {E.}~\bibnamefont
  {Pluharova}}, \bibinfo {author} {\bibfnamefont {M.~D.}\ \bibnamefont {Baer}},
  \bibinfo {author} {\bibfnamefont {G.~K.}\ \bibnamefont {Schenter}}, \bibinfo
  {author} {\bibfnamefont {P.}~\bibnamefont {Jungwirth}}, \ and\ \bibinfo
  {author} {\bibfnamefont {C.~J.}\ \bibnamefont {Mundy}},\ }\href@noop {}
  {\bibfield  {journal} {\bibinfo  {journal} {The Journal of Physical Chemistry
  B}\ }\textbf {\bibinfo {volume} {120}},\ \bibinfo {pages} {1749} (\bibinfo
  {year} {2016})}\BibitemShut {NoStop}%
\bibitem [{\citenamefont {Roy}\ \emph {et~al.}(2016)\citenamefont {Roy},
  \citenamefont {Baer}, \citenamefont {Mundy},\ and\ \citenamefont
  {Schenter}}]{Roy2016}%
  \BibitemOpen
  \bibfield  {author} {\bibinfo {author} {\bibfnamefont {S.}~\bibnamefont
  {Roy}}, \bibinfo {author} {\bibfnamefont {M.~D.}\ \bibnamefont {Baer}},
  \bibinfo {author} {\bibfnamefont {C.~J.}\ \bibnamefont {Mundy}}, \ and\
  \bibinfo {author} {\bibfnamefont {G.~K.}\ \bibnamefont {Schenter}},\
  }\href@noop {} {\bibfield  {journal} {\bibinfo  {journal} {The Journal of
  Physical Chemistry C}\ }\textbf {\bibinfo {volume} {120}},\ \bibinfo {pages}
  {7597} (\bibinfo {year} {2016})}\BibitemShut {NoStop}%
\bibitem [{\citenamefont {Dellago}\ \emph {et~al.}(1998)\citenamefont
  {Dellago}, \citenamefont {Bolhuis}, \citenamefont {Csajka},\ and\
  \citenamefont {Chandler}}]{TPS}%
  \BibitemOpen
  \bibfield  {author} {\bibinfo {author} {\bibfnamefont {C.}~\bibnamefont
  {Dellago}}, \bibinfo {author} {\bibfnamefont {P.~G.}\ \bibnamefont
  {Bolhuis}}, \bibinfo {author} {\bibfnamefont {F.~S.}\ \bibnamefont {Csajka}},
  \ and\ \bibinfo {author} {\bibfnamefont {D.}~\bibnamefont {Chandler}},\
  }\href@noop {} {\bibfield  {journal} {\bibinfo  {journal} {The Journal of
  chemical physics}\ }\textbf {\bibinfo {volume} {108}},\ \bibinfo {pages}
  {1964} (\bibinfo {year} {1998})}\BibitemShut {NoStop}%
\bibitem [{\citenamefont {Best}\ and\ \citenamefont
  {Hummer}(2005{\natexlab{a}})}]{HummerPNAS2005}%
  \BibitemOpen
  \bibfield  {author} {\bibinfo {author} {\bibfnamefont {R.~B.}\ \bibnamefont
  {Best}}\ and\ \bibinfo {author} {\bibfnamefont {G.}~\bibnamefont {Hummer}},\
  }\href
  {https://www.ncbi.nlm.nih.gov/pmc/articles/PMC1100744/pdf/pnas-0408098102.pdf}
  {\bibfield  {journal} {\bibinfo  {journal} {Proceedings of the National
  Academy of Sciences}\ }\textbf {\bibinfo {volume} {102}},\ \bibinfo {pages}
  {6732} (\bibinfo {year} {2005}{\natexlab{a}})}\BibitemShut {NoStop}%
\bibitem [{\citenamefont {Ma}\ and\ \citenamefont {Dinner}(2005)}]{Ma2005}%
  \BibitemOpen
  \bibfield  {author} {\bibinfo {author} {\bibfnamefont {A.}~\bibnamefont
  {Ma}}\ and\ \bibinfo {author} {\bibfnamefont {A.~R.}\ \bibnamefont
  {Dinner}},\ }\href@noop {} {\bibfield  {journal} {\bibinfo  {journal} {The
  Journal of Physical Chemistry B}\ }\textbf {\bibinfo {volume} {109}},\
  \bibinfo {pages} {6769} (\bibinfo {year} {2005})}\BibitemShut {NoStop}%
\bibitem [{\citenamefont {Peters}\ and\ \citenamefont
  {Trout}(2006)}]{peters2006obtaining}%
  \BibitemOpen
  \bibfield  {author} {\bibinfo {author} {\bibfnamefont {B.}~\bibnamefont
  {Peters}}\ and\ \bibinfo {author} {\bibfnamefont {B.~L.}\ \bibnamefont
  {Trout}},\ }\href@noop {} {\bibfield  {journal} {\bibinfo  {journal} {The
  Journal of chemical physics}\ }\textbf {\bibinfo {volume} {125}},\ \bibinfo
  {pages} {054108} (\bibinfo {year} {2006})}\BibitemShut {NoStop}%
\bibitem [{\citenamefont {Peters}(2016)}]{peters2016}%
  \BibitemOpen
  \bibfield  {author} {\bibinfo {author} {\bibfnamefont {B.}~\bibnamefont
  {Peters}},\ }\href@noop {} {\bibfield  {journal} {\bibinfo  {journal} {Annual
  review of physical chemistry}\ }\textbf {\bibinfo {volume} {67}},\ \bibinfo
  {pages} {669} (\bibinfo {year} {2016})}\BibitemShut {NoStop}%
\bibitem [{\citenamefont {Nadler}\ \emph {et~al.}(2006)\citenamefont {Nadler},
  \citenamefont {Lafon}, \citenamefont {Coifman},\ and\ \citenamefont
  {Kevrekidis}}]{diffusion_map_and_RC}%
  \BibitemOpen
  \bibfield  {author} {\bibinfo {author} {\bibfnamefont {B.}~\bibnamefont
  {Nadler}}, \bibinfo {author} {\bibfnamefont {S.}~\bibnamefont {Lafon}},
  \bibinfo {author} {\bibfnamefont {R.~R.}\ \bibnamefont {Coifman}}, \ and\
  \bibinfo {author} {\bibfnamefont {I.~G.}\ \bibnamefont {Kevrekidis}},\
  }\href@noop {} {\bibfield  {journal} {\bibinfo  {journal} {Applied and
  Computational Harmonic Analysis}\ }\textbf {\bibinfo {volume} {21}},\
  \bibinfo {pages} {113} (\bibinfo {year} {2006})}\BibitemShut {NoStop}%
\bibitem [{\citenamefont {Coifman}\ \emph {et~al.}(2008)\citenamefont
  {Coifman}, \citenamefont {Kevrekidis}, \citenamefont {Lafon}, \citenamefont
  {Maggioni},\ and\ \citenamefont {Nadler}}]{Coifman2008}%
  \BibitemOpen
  \bibfield  {author} {\bibinfo {author} {\bibfnamefont {R.~R.}\ \bibnamefont
  {Coifman}}, \bibinfo {author} {\bibfnamefont {I.~G.}\ \bibnamefont
  {Kevrekidis}}, \bibinfo {author} {\bibfnamefont {S.}~\bibnamefont {Lafon}},
  \bibinfo {author} {\bibfnamefont {M.}~\bibnamefont {Maggioni}}, \ and\
  \bibinfo {author} {\bibfnamefont {B.}~\bibnamefont {Nadler}},\ }\href@noop {}
  {\bibfield  {journal} {\bibinfo  {journal} {Multiscale Modeling \&
  Simulation}\ }\textbf {\bibinfo {volume} {7}},\ \bibinfo {pages} {842}
  (\bibinfo {year} {2008})}\BibitemShut {NoStop}%
\bibitem [{\citenamefont {Rohrdanz}\ \emph {et~al.}(2011)\citenamefont
  {Rohrdanz}, \citenamefont {Zheng}, \citenamefont {Maggioni},\ and\
  \citenamefont {Clementi}}]{Rohrdanz2011}%
  \BibitemOpen
  \bibfield  {author} {\bibinfo {author} {\bibfnamefont {M.~A.}\ \bibnamefont
  {Rohrdanz}}, \bibinfo {author} {\bibfnamefont {W.}~\bibnamefont {Zheng}},
  \bibinfo {author} {\bibfnamefont {M.}~\bibnamefont {Maggioni}}, \ and\
  \bibinfo {author} {\bibfnamefont {C.}~\bibnamefont {Clementi}},\ }\href@noop
  {} {\bibfield  {journal} {\bibinfo  {journal} {The Journal of chemical
  physics}\ }\textbf {\bibinfo {volume} {134}},\ \bibinfo {pages} {03B624}
  (\bibinfo {year} {2011})}\BibitemShut {NoStop}%
\bibitem [{\citenamefont {Noé}\ and\ \citenamefont {Nuske}(2013)}]{VAC}%
  \BibitemOpen
  \bibfield  {author} {\bibinfo {author} {\bibfnamefont {F.}~\bibnamefont
  {Noé}}\ and\ \bibinfo {author} {\bibfnamefont {F.}~\bibnamefont {Nuske}},\
  }\href@noop {} {\bibfield  {journal} {\bibinfo  {journal} {Multiscale
  Modeling \& Simulation}\ }\textbf {\bibinfo {volume} {11}},\ \bibinfo {pages}
  {635} (\bibinfo {year} {2013})}\BibitemShut {NoStop}%
\bibitem [{\citenamefont {P{\'e}rez-Hern{\'a}ndez}\ \emph
  {et~al.}(2013)\citenamefont {P{\'e}rez-Hern{\'a}ndez}, \citenamefont {Paul},
  \citenamefont {Giorgino}, \citenamefont {De~Fabritiis},\ and\ \citenamefont
  {No{\'e}}}]{TICA}%
  \BibitemOpen
  \bibfield  {author} {\bibinfo {author} {\bibfnamefont {G.}~\bibnamefont
  {P{\'e}rez-Hern{\'a}ndez}}, \bibinfo {author} {\bibfnamefont
  {F.}~\bibnamefont {Paul}}, \bibinfo {author} {\bibfnamefont {T.}~\bibnamefont
  {Giorgino}}, \bibinfo {author} {\bibfnamefont {G.}~\bibnamefont
  {De~Fabritiis}}, \ and\ \bibinfo {author} {\bibfnamefont {F.}~\bibnamefont
  {No{\'e}}},\ }\href@noop {} {\bibfield  {journal} {\bibinfo  {journal} {The
  Journal of chemical physics}\ }\textbf {\bibinfo {volume} {139}},\ \bibinfo
  {pages} {07B604\_1} (\bibinfo {year} {2013})}\BibitemShut {NoStop}%
\bibitem [{\citenamefont {Mardt}\ \emph {et~al.}(2018)\citenamefont {Mardt},
  \citenamefont {Pasquali}, \citenamefont {Wu},\ and\ \citenamefont
  {Noé}}]{VAMPnets}%
  \BibitemOpen
  \bibfield  {author} {\bibinfo {author} {\bibfnamefont {A.}~\bibnamefont
  {Mardt}}, \bibinfo {author} {\bibfnamefont {L.}~\bibnamefont {Pasquali}},
  \bibinfo {author} {\bibfnamefont {H.}~\bibnamefont {Wu}}, \ and\ \bibinfo
  {author} {\bibfnamefont {F.}~\bibnamefont {Noé}},\ }\href@noop {} {\bibfield
   {journal} {\bibinfo  {journal} {Nature communications}\ }\textbf {\bibinfo
  {volume} {9}},\ \bibinfo {pages} {1} (\bibinfo {year} {2018})}\BibitemShut
  {NoStop}%
\bibitem [{\citenamefont {Tiwary}\ and\ \citenamefont {Berne}(2016)}]{SGOOP}%
  \BibitemOpen
  \bibfield  {author} {\bibinfo {author} {\bibfnamefont {P.}~\bibnamefont
  {Tiwary}}\ and\ \bibinfo {author} {\bibfnamefont {B.}~\bibnamefont {Berne}},\
  }\href@noop {} {\bibfield  {journal} {\bibinfo  {journal} {Proceedings of the
  National Academy of Sciences}\ }\textbf {\bibinfo {volume} {113}},\ \bibinfo
  {pages} {2839} (\bibinfo {year} {2016})}\BibitemShut {NoStop}%
\bibitem [{\citenamefont {Hernandez}\ \emph {et~al.}(2018)\citenamefont
  {Hernandez}, \citenamefont {Wayment-Steele}, \citenamefont {Sultan},
  \citenamefont {Husic},\ and\ \citenamefont {Pande}}]{VDE}%
  \BibitemOpen
  \bibfield  {author} {\bibinfo {author} {\bibfnamefont {C.~X.}\ \bibnamefont
  {Hernandez}}, \bibinfo {author} {\bibfnamefont {H.~K.}\ \bibnamefont
  {Wayment-Steele}}, \bibinfo {author} {\bibfnamefont {M.~M.}\ \bibnamefont
  {Sultan}}, \bibinfo {author} {\bibfnamefont {B.~E.}\ \bibnamefont {Husic}}, \
  and\ \bibinfo {author} {\bibfnamefont {V.~S.}\ \bibnamefont {Pande}},\ }\href
  {\doibase 10.1103/PhysRevE.97.062412} {\bibfield  {journal} {\bibinfo
  {journal} {Phys Rev E}\ }\textbf {\bibinfo {volume} {97}},\ \bibinfo {pages}
  {062412} (\bibinfo {year} {2018})}\BibitemShut {NoStop}%
\bibitem [{\citenamefont {Ribeiro}\ \emph {et~al.}(2018)\citenamefont
  {Ribeiro}, \citenamefont {Bravo}, \citenamefont {Wang},\ and\ \citenamefont
  {Tiwary}}]{RAVE}%
  \BibitemOpen
  \bibfield  {author} {\bibinfo {author} {\bibfnamefont {J.~M.~L.}\
  \bibnamefont {Ribeiro}}, \bibinfo {author} {\bibfnamefont {P.}~\bibnamefont
  {Bravo}}, \bibinfo {author} {\bibfnamefont {Y.}~\bibnamefont {Wang}}, \ and\
  \bibinfo {author} {\bibfnamefont {P.}~\bibnamefont {Tiwary}},\ }\href
  {https://aip.scitation.org/doi/10.1063/1.5025487} {\bibfield  {journal}
  {\bibinfo  {journal} {The Journal of Chemical Physics}\ }\textbf {\bibinfo
  {volume} {149}},\ \bibinfo {pages} {072301} (\bibinfo {year}
  {2018})}\BibitemShut {NoStop}%
\bibitem [{\citenamefont {Wang}, \citenamefont {Ribeiro},\ and\ \citenamefont
  {Tiwary}(2019)}]{pRAVE}%
  \BibitemOpen
  \bibfield  {author} {\bibinfo {author} {\bibfnamefont {Y.}~\bibnamefont
  {Wang}}, \bibinfo {author} {\bibfnamefont {J.~M.~L.}\ \bibnamefont
  {Ribeiro}}, \ and\ \bibinfo {author} {\bibfnamefont {P.}~\bibnamefont
  {Tiwary}},\ }\href {\doibase 10.1038/s41467-019-11405-4} {\bibfield
  {journal} {\bibinfo  {journal} {Nature communications}\ }\textbf {\bibinfo
  {volume} {10}},\ \bibinfo {pages} {3573} (\bibinfo {year}
  {2019})}\BibitemShut {NoStop}%
\bibitem [{\citenamefont {Kingma}\ and\ \citenamefont {Welling}(2013)}]{VAE}%
  \BibitemOpen
  \bibfield  {author} {\bibinfo {author} {\bibfnamefont {D.~P.}\ \bibnamefont
  {Kingma}}\ and\ \bibinfo {author} {\bibfnamefont {M.}~\bibnamefont
  {Welling}},\ }\href@noop {} {\bibfield  {journal} {\bibinfo  {journal} {arXiv
  preprint arXiv:1312.6114}\ } (\bibinfo {year} {2013})}\BibitemShut {NoStop}%
\bibitem [{\citenamefont {Alemi}\ \emph {et~al.}(2016)\citenamefont {Alemi},
  \citenamefont {Fischer}, \citenamefont {Dillon},\ and\ \citenamefont
  {Murphy}}]{variational_IB}%
  \BibitemOpen
  \bibfield  {author} {\bibinfo {author} {\bibfnamefont {A.~A.}\ \bibnamefont
  {Alemi}}, \bibinfo {author} {\bibfnamefont {I.}~\bibnamefont {Fischer}},
  \bibinfo {author} {\bibfnamefont {J.~V.}\ \bibnamefont {Dillon}}, \ and\
  \bibinfo {author} {\bibfnamefont {K.}~\bibnamefont {Murphy}},\ }\href@noop {}
  {\bibfield  {journal} {\bibinfo  {journal} {arXiv preprint arXiv:1612.00410}\
  } (\bibinfo {year} {2016})}\BibitemShut {NoStop}%
\bibitem [{\citenamefont {Tishby}, \citenamefont {Pereira},\ and\ \citenamefont
  {Bialek}(2000)}]{IB}%
  \BibitemOpen
  \bibfield  {author} {\bibinfo {author} {\bibfnamefont {N.}~\bibnamefont
  {Tishby}}, \bibinfo {author} {\bibfnamefont {F.~C.}\ \bibnamefont {Pereira}},
  \ and\ \bibinfo {author} {\bibfnamefont {W.}~\bibnamefont {Bialek}},\
  }\href@noop {} {\bibfield  {journal} {\bibinfo  {journal} {arXiv preprint
  physics/0004057}\ } (\bibinfo {year} {2000})}\BibitemShut {NoStop}%
\bibitem [{\citenamefont {Berlekamp}(2015)}]{coding_theory}%
  \BibitemOpen
  \bibfield  {author} {\bibinfo {author} {\bibfnamefont {E.~R.}\ \bibnamefont
  {Berlekamp}},\ }\href@noop {} {\emph {\bibinfo {title} {Algebraic coding
  theory (revised edition)}}}\ (\bibinfo  {publisher} {World Scientific},\
  \bibinfo {year} {2015})\BibitemShut {NoStop}%
\bibitem [{\citenamefont {Shannon}(1959)}]{RD_theory}%
  \BibitemOpen
  \bibfield  {author} {\bibinfo {author} {\bibfnamefont {C.~E.}\ \bibnamefont
  {Shannon}},\ }\href@noop {} {\bibfield  {journal} {\bibinfo  {journal} {IRE
  Nat. Conv. Rec}\ }\textbf {\bibinfo {volume} {4}},\ \bibinfo {pages} {1}
  (\bibinfo {year} {1959})}\BibitemShut {NoStop}%
\bibitem [{\citenamefont {Alemi}\ \emph {et~al.}(2018)\citenamefont {Alemi},
  \citenamefont {Poole}, \citenamefont {Fischer}, \citenamefont {Dillon},
  \citenamefont {Saurus},\ and\ \citenamefont {Murphy}}]{alemi2018an}%
  \BibitemOpen
  \bibfield  {author} {\bibinfo {author} {\bibfnamefont {A.}~\bibnamefont
  {Alemi}}, \bibinfo {author} {\bibfnamefont {B.}~\bibnamefont {Poole}},
  \bibinfo {author} {\bibfnamefont {I.}~\bibnamefont {Fischer}}, \bibinfo
  {author} {\bibfnamefont {J.}~\bibnamefont {Dillon}}, \bibinfo {author}
  {\bibfnamefont {R.~A.}\ \bibnamefont {Saurus}}, \ and\ \bibinfo {author}
  {\bibfnamefont {K.}~\bibnamefont {Murphy}},\ }\href
  {https://openreview.net/forum?id=H1rRWl-Cb} {\  (\bibinfo {year}
  {2018})}\BibitemShut {NoStop}%
\bibitem [{\citenamefont {Alemi}\ and\ \citenamefont
  {Fischer}(2018)}]{alemi2018therml}%
  \BibitemOpen
  \bibfield  {author} {\bibinfo {author} {\bibfnamefont {A.~A.}\ \bibnamefont
  {Alemi}}\ and\ \bibinfo {author} {\bibfnamefont {I.}~\bibnamefont
  {Fischer}},\ }\href@noop {} {\bibfield  {journal} {\bibinfo  {journal} {arXiv
  preprint arXiv:1807.04162}\ } (\bibinfo {year} {2018})}\BibitemShut {NoStop}%
\bibitem [{\citenamefont {Best}\ and\ \citenamefont
  {Hummer}(2005{\natexlab{b}})}]{best2005reaction}%
  \BibitemOpen
  \bibfield  {author} {\bibinfo {author} {\bibfnamefont {R.~B.}\ \bibnamefont
  {Best}}\ and\ \bibinfo {author} {\bibfnamefont {G.}~\bibnamefont {Hummer}},\
  }\href@noop {} {\bibfield  {journal} {\bibinfo  {journal} {Proceedings of the
  National Academy of Sciences}\ }\textbf {\bibinfo {volume} {102}},\ \bibinfo
  {pages} {6732} (\bibinfo {year} {2005}{\natexlab{b}})}\BibitemShut {NoStop}%
\bibitem [{\citenamefont {Tomczak}\ and\ \citenamefont
  {Welling}(2017)}]{VampPrior}%
  \BibitemOpen
  \bibfield  {author} {\bibinfo {author} {\bibfnamefont {J.~M.}\ \bibnamefont
  {Tomczak}}\ and\ \bibinfo {author} {\bibfnamefont {M.}~\bibnamefont
  {Welling}},\ }\href@noop {} {\bibfield  {journal} {\bibinfo  {journal} {arXiv
  preprint arXiv:1705.07120}\ } (\bibinfo {year} {2017})}\BibitemShut {NoStop}%
\bibitem [{\citenamefont {Berezansky}, \citenamefont {Sheftel},\ and\
  \citenamefont {Us}(1996)}]{Spectral_Decomp}%
  \BibitemOpen
  \bibfield  {author} {\bibinfo {author} {\bibfnamefont {Y.~M.}\ \bibnamefont
  {Berezansky}}, \bibinfo {author} {\bibfnamefont {Z.~G.}\ \bibnamefont
  {Sheftel}}, \ and\ \bibinfo {author} {\bibfnamefont {G.~F.}\ \bibnamefont
  {Us}},\ }\enquote {\bibinfo {title} {Spectral decomposition of compact
  selfadjoint operators. analytic functions of operators},}\ in\ \href@noop {}
  {\emph {\bibinfo {booktitle} {Functional Analysis}}}\ (\bibinfo  {publisher}
  {Springer},\ \bibinfo {year} {1996})\ pp.\ \bibinfo {pages}
  {355--384}\BibitemShut {NoStop}%
\bibitem [{\citenamefont {Swope}, \citenamefont {Pitera},\ and\ \citenamefont
  {Suits}(2004)}]{Markov_model}%
  \BibitemOpen
  \bibfield  {author} {\bibinfo {author} {\bibfnamefont {W.~C.}\ \bibnamefont
  {Swope}}, \bibinfo {author} {\bibfnamefont {J.~W.}\ \bibnamefont {Pitera}}, \
  and\ \bibinfo {author} {\bibfnamefont {F.}~\bibnamefont {Suits}},\
  }\href@noop {} {\bibfield  {journal} {\bibinfo  {journal} {The Journal of
  Physical Chemistry B}\ }\textbf {\bibinfo {volume} {108}},\ \bibinfo {pages}
  {6571} (\bibinfo {year} {2004})}\BibitemShut {NoStop}%
\bibitem [{\citenamefont {Noé}\ and\ \citenamefont
  {Fischer}(2008)}]{transition_networks}%
  \BibitemOpen
  \bibfield  {author} {\bibinfo {author} {\bibfnamefont {F.}~\bibnamefont
  {Noé}}\ and\ \bibinfo {author} {\bibfnamefont {S.}~\bibnamefont {Fischer}},\
  }\href@noop {} {\bibfield  {journal} {\bibinfo  {journal} {Current opinion in
  structural biology}\ }\textbf {\bibinfo {volume} {18}},\ \bibinfo {pages}
  {154} (\bibinfo {year} {2008})}\BibitemShut {NoStop}%
\bibitem [{\citenamefont {Wales}(2009)}]{Wales2009}%
  \BibitemOpen
  \bibfield  {author} {\bibinfo {author} {\bibfnamefont {D.~J.}\ \bibnamefont
  {Wales}},\ }\href@noop {} {\bibfield  {journal} {\bibinfo  {journal} {The
  Journal of chemical physics}\ }\textbf {\bibinfo {volume} {130}},\ \bibinfo
  {pages} {204111} (\bibinfo {year} {2009})}\BibitemShut {NoStop}%
\bibitem [{\citenamefont {Lane}\ \emph {et~al.}(2011)\citenamefont {Lane},
  \citenamefont {Bowman}, \citenamefont {Beauchamp}, \citenamefont {Voelz},\
  and\ \citenamefont {Pande}}]{Lane2011}%
  \BibitemOpen
  \bibfield  {author} {\bibinfo {author} {\bibfnamefont {T.~J.}\ \bibnamefont
  {Lane}}, \bibinfo {author} {\bibfnamefont {G.~R.}\ \bibnamefont {Bowman}},
  \bibinfo {author} {\bibfnamefont {K.}~\bibnamefont {Beauchamp}}, \bibinfo
  {author} {\bibfnamefont {V.~A.}\ \bibnamefont {Voelz}}, \ and\ \bibinfo
  {author} {\bibfnamefont {V.~S.}\ \bibnamefont {Pande}},\ }\href@noop {}
  {\bibfield  {journal} {\bibinfo  {journal} {Journal of the American Chemical
  Society}\ }\textbf {\bibinfo {volume} {133}},\ \bibinfo {pages} {18413}
  (\bibinfo {year} {2011})}\BibitemShut {NoStop}%
\bibitem [{\citenamefont {Thiede}\ \emph {et~al.}(2019)\citenamefont {Thiede},
  \citenamefont {Giannakis}, \citenamefont {Dinner},\ and\ \citenamefont
  {Weare}}]{Galerkin_approximation}%
  \BibitemOpen
  \bibfield  {author} {\bibinfo {author} {\bibfnamefont {E.~H.}\ \bibnamefont
  {Thiede}}, \bibinfo {author} {\bibfnamefont {D.}~\bibnamefont {Giannakis}},
  \bibinfo {author} {\bibfnamefont {A.~R.}\ \bibnamefont {Dinner}}, \ and\
  \bibinfo {author} {\bibfnamefont {J.}~\bibnamefont {Weare}},\ }\href@noop {}
  {\bibfield  {journal} {\bibinfo  {journal} {The Journal of chemical physics}\
  }\textbf {\bibinfo {volume} {150}},\ \bibinfo {pages} {244111} (\bibinfo
  {year} {2019})}\BibitemShut {NoStop}%
\bibitem [{\citenamefont {Biswas}, \citenamefont {Lickert},\ and\ \citenamefont
  {Stock}(2018)}]{biswas2018metadynamics}%
  \BibitemOpen
  \bibfield  {author} {\bibinfo {author} {\bibfnamefont {M.}~\bibnamefont
  {Biswas}}, \bibinfo {author} {\bibfnamefont {B.}~\bibnamefont {Lickert}}, \
  and\ \bibinfo {author} {\bibfnamefont {G.}~\bibnamefont {Stock}},\
  }\href@noop {} {\bibfield  {journal} {\bibinfo  {journal} {The Journal of
  Physical Chemistry B}\ }\textbf {\bibinfo {volume} {122}},\ \bibinfo {pages}
  {5508} (\bibinfo {year} {2018})}\BibitemShut {NoStop}%
\bibitem [{\citenamefont {Bussi}\ and\ \citenamefont
  {Parrinello}(2007)}]{Langevin}%
  \BibitemOpen
  \bibfield  {author} {\bibinfo {author} {\bibfnamefont {G.}~\bibnamefont
  {Bussi}}\ and\ \bibinfo {author} {\bibfnamefont {M.}~\bibnamefont
  {Parrinello}},\ }\href@noop {} {\bibfield  {journal} {\bibinfo  {journal}
  {Physical Review E}\ }\textbf {\bibinfo {volume} {75}},\ \bibinfo {pages}
  {056707} (\bibinfo {year} {2007})}\BibitemShut {NoStop}%
\bibitem [{\citenamefont {Berendsen}, \citenamefont {van~der Spoel},\ and\
  \citenamefont {van Drunen}(1995)}]{GROMACS1995}%
  \BibitemOpen
  \bibfield  {author} {\bibinfo {author} {\bibfnamefont {H.~J.}\ \bibnamefont
  {Berendsen}}, \bibinfo {author} {\bibfnamefont {D.}~\bibnamefont {van~der
  Spoel}}, \ and\ \bibinfo {author} {\bibfnamefont {R.}~\bibnamefont {van
  Drunen}},\ }\href@noop {} {\bibfield  {journal} {\bibinfo  {journal}
  {Computer physics communications}\ }\textbf {\bibinfo {volume} {91}},\
  \bibinfo {pages} {43} (\bibinfo {year} {1995})}\BibitemShut {NoStop}%
\bibitem [{\citenamefont {Abraham}\ \emph {et~al.}(2015)\citenamefont
  {Abraham}, \citenamefont {Murtola}, \citenamefont {Schulz}, \citenamefont
  {Páll}, \citenamefont {Smith}, \citenamefont {Hess},\ and\ \citenamefont
  {Lindahl}}]{GROMACS2015}%
  \BibitemOpen
  \bibfield  {author} {\bibinfo {author} {\bibfnamefont {M.~J.}\ \bibnamefont
  {Abraham}}, \bibinfo {author} {\bibfnamefont {T.}~\bibnamefont {Murtola}},
  \bibinfo {author} {\bibfnamefont {R.}~\bibnamefont {Schulz}}, \bibinfo
  {author} {\bibfnamefont {S.}~\bibnamefont {Páll}}, \bibinfo {author}
  {\bibfnamefont {J.~C.}\ \bibnamefont {Smith}}, \bibinfo {author}
  {\bibfnamefont {B.}~\bibnamefont {Hess}}, \ and\ \bibinfo {author}
  {\bibfnamefont {E.}~\bibnamefont {Lindahl}},\ }\href@noop {} {\bibfield
  {journal} {\bibinfo  {journal} {SoftwareX}\ }\textbf {\bibinfo {volume}
  {1}},\ \bibinfo {pages} {19} (\bibinfo {year} {2015})}\BibitemShut {NoStop}%
\bibitem [{\citenamefont {Tribello}\ \emph {et~al.}(2014)\citenamefont
  {Tribello}, \citenamefont {Bonomi}, \citenamefont {Branduardi}, \citenamefont
  {Camilloni},\ and\ \citenamefont {Bussi}}]{PLUMED2}%
  \BibitemOpen
  \bibfield  {author} {\bibinfo {author} {\bibfnamefont {G.~A.}\ \bibnamefont
  {Tribello}}, \bibinfo {author} {\bibfnamefont {M.}~\bibnamefont {Bonomi}},
  \bibinfo {author} {\bibfnamefont {D.}~\bibnamefont {Branduardi}}, \bibinfo
  {author} {\bibfnamefont {C.}~\bibnamefont {Camilloni}}, \ and\ \bibinfo
  {author} {\bibfnamefont {G.}~\bibnamefont {Bussi}},\ }\href@noop {}
  {\bibfield  {journal} {\bibinfo  {journal} {Computer Physics Communications}\
  }\textbf {\bibinfo {volume} {185}},\ \bibinfo {pages} {604} (\bibinfo {year}
  {2014})}\BibitemShut {NoStop}%
\bibitem [{\citenamefont {Bussi}, \citenamefont {Donadio},\ and\ \citenamefont
  {Parrinello}(2007)}]{v_rescaling}%
  \BibitemOpen
  \bibfield  {author} {\bibinfo {author} {\bibfnamefont {G.}~\bibnamefont
  {Bussi}}, \bibinfo {author} {\bibfnamefont {D.}~\bibnamefont {Donadio}}, \
  and\ \bibinfo {author} {\bibfnamefont {M.}~\bibnamefont {Parrinello}},\
  }\href@noop {} {\bibfield  {journal} {\bibinfo  {journal} {The Journal of
  chemical physics}\ }\textbf {\bibinfo {volume} {126}},\ \bibinfo {pages}
  {014101} (\bibinfo {year} {2007})}\BibitemShut {NoStop}%
\bibitem [{\citenamefont {Kingma}\ and\ \citenamefont {Ba}(2014)}]{Adam}%
  \BibitemOpen
  \bibfield  {author} {\bibinfo {author} {\bibfnamefont {D.~P.}\ \bibnamefont
  {Kingma}}\ and\ \bibinfo {author} {\bibfnamefont {J.}~\bibnamefont {Ba}},\
  }\href@noop {} {\bibfield  {journal} {\bibinfo  {journal} {arXiv preprint
  arXiv:1412.6980}\ } (\bibinfo {year} {2014})}\BibitemShut {NoStop}%
\bibitem [{\citenamefont {Mori}\ \emph {et~al.}(2020)\citenamefont {Mori},
  \citenamefont {Okazaki}, \citenamefont {Mori}, \citenamefont {Kim},\ and\
  \citenamefont {Matubayasi}}]{Mori2020}%
  \BibitemOpen
  \bibfield  {author} {\bibinfo {author} {\bibfnamefont {Y.}~\bibnamefont
  {Mori}}, \bibinfo {author} {\bibfnamefont {K.-i.}\ \bibnamefont {Okazaki}},
  \bibinfo {author} {\bibfnamefont {T.}~\bibnamefont {Mori}}, \bibinfo {author}
  {\bibfnamefont {K.}~\bibnamefont {Kim}}, \ and\ \bibinfo {author}
  {\bibfnamefont {N.}~\bibnamefont {Matubayasi}},\ }\href@noop {} {\bibfield
  {journal} {\bibinfo  {journal} {arXiv preprint arXiv:.13186}\ } (\bibinfo
  {year} {2020})}\BibitemShut {NoStop}%
\bibitem [{\citenamefont {Bussi}\ and\ \citenamefont {Laio}(2020)}]{Bussi2020}%
  \BibitemOpen
  \bibfield  {author} {\bibinfo {author} {\bibfnamefont {G.}~\bibnamefont
  {Bussi}}\ and\ \bibinfo {author} {\bibfnamefont {A.}~\bibnamefont {Laio}},\
  }\href@noop {} {\bibfield  {journal} {\bibinfo  {journal} {Nature Reviews
  Physics}\ ,\ \bibinfo {pages} {1}} (\bibinfo {year} {2020})}\BibitemShut
  {NoStop}%
\bibitem [{\citenamefont {Bengio}, \citenamefont {Courville},\ and\
  \citenamefont {Vincent}(2013)}]{Bengio2013}%
  \BibitemOpen
  \bibfield  {author} {\bibinfo {author} {\bibfnamefont {Y.}~\bibnamefont
  {Bengio}}, \bibinfo {author} {\bibfnamefont {A.}~\bibnamefont {Courville}}, \
  and\ \bibinfo {author} {\bibfnamefont {P.}~\bibnamefont {Vincent}},\
  }\href@noop {} {\bibfield  {journal} {\bibinfo  {journal} {IEEE transactions
  on pattern analysis and machine intelligence}\ }\textbf {\bibinfo {volume}
  {35}},\ \bibinfo {pages} {1798} (\bibinfo {year} {2013})}\BibitemShut
  {NoStop}%
\end{thebibliography}%

\end{document}